\documentstyle[aps,epsfig,multicol,eqsecnum]{revtex}    
\draft

\begin{document}
\title{Optical response of supported particles} 
\author{Ingve Simonsen$^{1,2,}$\footnote{Email : Ingve.Simonsen@phys.ntnu.no}, 
        R\'emi Lazzari$^{1,}$\footnote{Email : Remi.Lazzari@sgr.saint-gobain.com},
        Jacques Jupille$^{1,}$\footnote{Email : Jacques.Jupille@sgr.saint-gobain.com}, 
        St\'ephane Roux$^{1,}$\footnote{Email : Stephane.Roux@sgr.saint-gobain.com} 
       }
\address{\vspace{0.5cm}
         $^1$Laboratoire Mixte CNRS/Saint-Gobain (UMR 125) ``Surface du Verre et Interfaces''\\
         39, Quai Lucien Lefranc, BP 135,\\
         F-93303 Aubervilliers Cedex,\\
         France\\
         \vspace{0.5cm}
         $^2$Department of Physics, Theoretical Physics Group, \\
         The Norwegian University of Science and Technology (NTNU), \\
         N-7034 Trondheim, \\
         Norway
         \vspace{0.3cm}
         }

\date{\today}
\maketitle


\begin{abstract}
    The present work reports a general method for the calculation of the polarizability of a truncated sphere on a substrate. A multipole expansion is used, where the multipoles are not necessarily localized in the center of the sphere but can freely move on the revolution axis. From the weak formulation of the boundary conditions, an infinite set of linear equations for the multipole coefficients is derived. To obtain this set, the interaction between the island and the substrate is taken into account by the technique of image multipoles.
For numerical implementation, this set is truncated at an arbitrary multipole order. The accuracy of the method is jugded through the stability of the truncated sphere polarizability and the the fulfilment of the boundary conditions which are demonstrated to be satisfied in large regions of the parameter space. This method brings an improvement with respect to the Bedeaux's case~\cite{Wind87a,Wind87b} where the multipoles are located in the center of the sphere.
\end{abstract}

 \vspace{5mm} 
 \pacs{PACS numbers:  }



\section{Introduction}

The {\it in situ} characterization of the growth mode of a thin film
in vacuum conditions is a long pending question. Indeed, a doubt is
always shed on the {\it ex situ} observations of deposits, often
performed by electron microscopy, which may not be representative of
the studied systems during their growth. Several diffraction
techniques are suited to examine the structure of the growing films.
For those many techniques probing surfaces by means of charged
species, such as low energy electron diffraction and reflection high
energy electron diffraction, the charge build-up prevents a
straightforward interpretation of the results in the case of
insulating substrates. Only electromagnetic probes or neutral atoms
diffraction can analyse most systems in a non disturbing way. However,
neutral atoms are only sensitive to the top most layer of a substrate
while, despite its impressive success in the field of surface science,
the grazing X-ray diffraction technique~\cite{Renaud_SSReports} is
limited in its application because it implies the use of high
intensity synchrotron radiations sources. Universally used in the
field of thin film growth, near field microscopies as atomic force and
scanning tunneling microscopies may imply a perturbation of the growth
and are hampered by the tip artefact. Moreover, the rather long time
needed to record an image with these methods often prevent them to be
used as a real time techniques. These limitations likely explain the
development of analytical methods based on optical probes for
monitoring the growth of a film and determining the thickness of the
deposited layers. Indeed, UV-visible tools fulfill the {\it in situ}
and non disturbing conditions for the examination of deposits during
their formation, combined with a simple and versatile use on most
substrates. In particular, optical methods, which have also the
capability of being run in any environment, are widely spread in
semiconductor technology~\cite{Aspnes}.

At the thermodynamical equilibrium, a film wetting the substrate on
which it is deposited is expected to grow layer by layer, following a
so-called Frank-Van der Merwe growth mode. On the contrary, when a
substrate is poorly wetted by the deposited material,
three-dimensional clusters are formed in a Volmer-Weber growth mode.
In an intermediate case, the Stransky-Krastanov growth mode, the
substrate is covered by a few wetting monolayers before the adlayer
relaxes to give rise to clusters formation. In addition, in many
cases, the morphology of deposited films not only depends on
thermodynamics, via the surface and interface energies, but also on
kinetics through energy barriers, diffusion coefficients and
intensities of the impiging fluxes. At temperatures and fluxes such
that the supersaturation is very high, growth of thin films can give
rise to the formation of clusters whose shapes are very far from those
expected in equilibrium conditions. Therefore, it is often very hard
to distinguish between the different growth modes at the earliest
stages of the formation of a film. The knowledge of the cluster mean
diameter, aspect ratio, and density during deposition would contribute
a breakthrough in the understanding of the growth. In this context,
surface differential reflectivity has been proved to be a powerful
tool for the determination of the shape ratio~\cite{Yves1,Doro1,Doro2}
and of the size of the metallic clusters~\cite{Doro3,Remi}, even
within the monolayer range. However, a quantitative analysis of the
optical spectra is still at its beginning~\cite{Remi}.

Since the pioneering work of Maxwell Garnett~\cite{Maxwell-Garnett} at
the turn of the century, there has been large scientific interest in
the optical properties of metallic clusters~\cite{KreibigBook}. Their
optical behaviours are driven to a large extent by the Mie
excitations~\cite{Mie} which can be viewed as surface
plasmon-polaritons. If, for isolated clusters with simple shapes, such
as spheres or spheroids in vacuum, the exact solution of the Maxwell
equation is known, the difficulty of a reliable description of
the optical properties of particles dramatically increases for
interacting particles of complex shapes, either in a matrix or on a
surface. Eventhough the Maxwell Garnett effective medium theory and
other such theories~\cite{Landauer} have been quite successful in
tackling these questions, an accurate description of the macroscopic
optical properties of the particles, as absorption and reflection,
requires a more sophisticated approach than the mean field theory to
find the renormalized polarizability which governs the far field
behaviour. Moreover, for clusters deposited on a surface, a
quantitative description of the optical properties of the thin film is
not only hampered by the interactions between aggregates but also by
the mutual interactions between the latter and the
substrate~\cite{Barrera2,Barrera1}.

The cluster-substrate and cluster-cluster interactions can be modelled
at the dipolar order as in the theory which has been developped by
Yamaguchi {\it et al}.~\cite{Yamaguchi1,Yamaguchi2}. The limitations
of such an approach, which does not allow for a quantitative
description of the size and shape ratio of the clusters of a
deposit~\cite{Doro2}, prompted Bedeaux and Vlieger~\cite{Bedeaux73} to
propose a theory, hereafter referred to as the Bedeaux-Vlieger model,
to account for the optical behaviour of a granular thin film
separating two bulk media. This was done by introducing some auxiliary
fields named excess fields which enabled to describe the macroscopic
optical effect of the film, with no need of a detailed description of
the spatial variations of these quantities as one moves away from
boundary layer~\cite{Bedeaux73}. The optical properties of the film is
essentially related to four surface susceptibilities~\cite{Bedeaux73}
which are nothing more than the total integrated excess field. These
surface susceptibilities which are linked to the island
polarizabilities for a discontinuous film govern the Fresnel optical
coefficients~\cite{Bedeaux74,Bedeaux76,Bedeaux80,Bedeaux83}. The
modelling of the dipoles arising in clusters under the light
excitation relies on the assumption that the mean size of clusters is
small compared to the wavelength of the light. Under these conditions,
retardation effects, which are related to the finiteness of the speed
of light, can be neglected around and inside the islands.  Hence, in
the electrostatic limit, the Maxwell equations for non-magnetic
materials and the Laplace equation for the electric potential describe
the same physics and are thus equivalent. Similar ideas had been put
forward earlier by Kretschmann~\cite{Kretschmann}, but using a
somewhat different formalism.

This method has been applied to truncated spheres on a substrate by
Wind {\it et al}.~\cite{Wind87a,Wind87b}. The Laplace equation is
solved by means of a multipole expansion technique~\cite{Jackson}
which consists in placing multipoles at the centre of the sphere. The
interaction with the substrate is taken into account by introducing
image multipoles located at the image point of the multipoles with
respect to the surface. The polarizability of the individual islands
is then renormalized by taking in account the interactions with the
neighbouring islands in a self-consistent
approximation~\cite{Bedeaux80}. This relatively simple model works
quite nicely under many circumstances~\cite{Remi,Bedeaux83,Bedeaux88}
and may reproduce the main experimental optical features to a large
extent (see Sec.~\ref{Sec:DifRef} below).

However, while trying to fit optical data corresponding
to disk-shaped clusters with aspect ratio higher than 2, which implies
a representation of the clusters by truncated spheres with centres
below the surface of the substrate~\cite{Doro2}, the numerical
computation turns out to be unstable. Beyond the necessity of a
general model that could be applied to any case, there appears a need
for clear criteria to establish the consistency of the calculation. In
the next section, the example of the optical response of a thin silver
film deposited on a magnesium oxide substrate allows to describe in
more details the position of the problem.

\section{Differential reflectivity on Ag/MgO(100) films}
\label{Sec:DifRef}
In a previous paper, it has been reported that UV-visible differential
reflectance data recorded during the deposit of a silver film on a
MgO(100) substrate in vacuum lead to estimates of the size, aspect
ratio and density of the clusters film which were in excellent
agreement with the values derived from an {\it ex situ} scanning
electron microscopy study of the same sample~\cite{Remi}.

\subsection{Experimental setup and results}
A silver film of average thickness $2$~nm was deposited on a
single-crystal MgO(100) held at $T=300^{\circ}C$ under ultra high
vacuum conditions by Knudsen evaporation.  During evaporation,
$p$-polarized light (i.e.\ with the electric field in the plane of
incidence) in the UV-visible energy range $1$--$5$~eV is impinged at
an incident angle of $\theta_0=45^\circ$. the reflected light is
recorded in the specular direction. The specular reflectivity,
$R(\omega)$, can be obtained as function of the frequency, $\omega$ of
the incident light. The recorded relevant quantity is the differential
reflection coefficient which is defined as 
\begin{eqnarray} 
  \label{delta_R_over_R}
  \frac{\Delta R(\omega)}{R(\omega)} 
        &=& \frac{R(\omega) -  R_F(\omega)}{R_F(\omega)}.
\end{eqnarray} 
where $R_F(\omega)$ is the Fresnel reflectivity of the bare substrate.
More detailed descriptions of the experimental setup and technique
used have been already published~\cite{Doro1,Remi}.

An ex situ image of the silver film has been collected by means of a field electron gun - scanning electron microscope (FEG-SEM). It is shown in Fig.~\ref{Fig:1}a). As can be seen from this image, the deposit
consists of small islands of a linear size of around $10$~nm
distributed over the surface of the substrate. Such a Volmer-Weber
growth mode is expected for the Ag/MgO system since noble metals
poorly wet wide band gap oxide surfaces~\cite{Fabrice}.

The experimental differential reflectivity spectra at the end of
growth is plotted in Fig.~\ref{Fig:1}b as function of the photon
energy, $E=\hbar \omega$, for an incident angle of $45^\circ$. The two
pronounced resonances seen in this graph are excited by the two
components of the $p$-polarized electric field. The low and
high-energy peaks are related to longitudinal and transversal plasma
oscillations inside the clusters, respectively~\cite{KreibigBook}.
Their positions in energy are mainly governed by the aspect-ratio of
the islands, which is given by the ratio of height over lateral size,
the electromagnetic coupling of the film with the substrate and the
interactions between particles~\cite{Doro1,Doro2,Doro3}.

\subsection{Limits of the simulation}
The solid curve in Fig.~\ref{Fig:1}b represents a simulation result
obtained by means of the model~\cite{Remi} derived from the method of
Wind {\it et al}~\cite{Wind87a,Wind87b}. Within the framework of this
theory, deposited islands are modelled by a set of identical truncated
spheres which are placed on a regular square array to simplify the
calculation since it has earlier been shown that, for low cluster
coverages, the optical response of a regular array of clusters with,
for example, a square or triangular lattice, marginally differs from
that of randomly distributed aggregates~\cite{Haarmans93}. Moreover,
for such a low cluster density, higher order interactions between
islands are expected to be negligible~\cite{Haarmans93}. The values
for the dielectric functions are taken from the
literature~\cite{Palik} with, in the case of silver, a correction to
account for the finite size of the clusters~\cite{Wind87b,KreibigBook}
which reduces the electron mean free path. The island polarizability
is evaluated in the quasistatic approximation, the use of which is
justified by the experimental energy range ($1$--$5$~eV, i.e.\ 
wavelength in the range $200$--$1200$~nm), by a multipolar development
of the potential that is truncated at an order $M$ for numerical
reasons. This polarizability is then renormalized by the inter-island
coupling which is accounted for only at dipolar order. The model
spectrum is chosen by means of a try and error method, the main
emphasis being put on trying to reproduce the location of the low- and
high-energy resonances and their intensity.

The theoretical result shown in Fig.~\ref{Fig:1}b corresponds to
islands represented by spheres of radius $R=6.8$~nm with the centre of
the sphere located at $0.11 R$ above the substrate. The islands are
placed on a square lattice with a inter-island distance (lattice
constant) of $19.6$~nm which corresponds to a cluster density of
$2.6\,10^{11}$~$\mbox{cm}^{-2}$. These values nicely compare to those
obtained by scanning electron microscopy (Fig.~\ref{Fig:1}a), a
cluster radius of $8.2\pm 1.5$~nm and a cluster density of $1.9\pm
0.5\,10^{11}$~$\mbox{cm}^{-2}$~\cite{Remi}. The silver coverage which
can be derived from the optics compares within few per cent with that
obtained from the SEM image. Such quantitative agreement is
impressive, in particular in view of the relative simplicity of the
model on which the simulation is based.

The best fit between the optical spectrum and the model has been
found~\cite{Remi} for an aspect ratio (diameter/height) of 1.8
(Fig.~\ref{Fig:1}b). Numerical results are shown in Fig.~\ref{Fig:2}
as a function of the multipole order within this particular geometry.
A good convergence is obtained upon increasing the multipole order.
The same is observed for any case in which the aspect ratio of the
deposited cluster is lower than 2~(i.e.\ when the center of the
sphere is above the surface of the substrate). However, when
performing the calculation for truncated spheres with centers below
the surface of the substrate, the numerical code does not converge any
longer. An example is given in Fig.~\ref{Fig:3} where the model is
applied to a cluster geometry similar to that used in Fig.~\ref{Fig:2}
except for the aspect ratio which is now set at a value of $2.2$. Upon
including higher and higher multipole orders in the simulations, the
position of the low energy resonance is wandering and shows no trend
of convergence. At this stage, two cases ought to be considered. When
the centre of the sphere is beyond the mid distance between the
surface of the substrate and the bottom of the sphere, the divergence
of the calculation happens for mathematical reasons since the image
point is outside the physical domain or, in other words, outside the
sphere. In a different way, when the image point is within the sphere,
solutions might exist. The lack of convergence likely arises from the
conditioning of the matrix. Indeed, the problem can not be solved by
increasing the number of multipoles since, at some stage, the machine
accuracy is overflowed. In both cases, a suggestion to find out
favourable configurations in which the numerical calculation could be
performed more successfully, is to move the expansion centre. This
point will be examined in Sect.~\ref{Sec:Model}.

Nevertheless, neither the convergence of the model itself nor that of
the model with the experimental data can guarantee the correctness of
the solution. These are only indirect proofs. In the model, potentials
are expressed as a function of expansion coefficients up to the $M$'th
order, although the optical response, which is derived from the dipole
polarizability of the deposited clusters, only relies on the first
order expansion coefficients. Therefore, the convergence of the
calculated optical response is necessary but not sufficient to
validate the expressions of the potentials. At variance, the boundary
conditions depend on all the expansion coefficients until the highest
order, so that a definite evidence of the consistency of the model
would be gained by their fulfilment. The question will be addressed in
Sect.~\ref{Sec:NumRes} where the above Ag/MgO case will be used to
test the accuracy of the model.


\section{A model for evaluating island polarizabilities}
\label{Sec:Model}

The cluster geometry used in the present paper, which is derived from
that used by Bedeaux {\it et al}. is depicted in Fig.~\ref{Fig:4}. It
consists of a substrate~(medium~2) localized in vacuum~(medium~1) and
covered with clusters~(medium~3) whose size is assumed to be small
with respect to the wavelength of the incident light. The islands are
modelled by truncated spheres of radius $R$. The vacuum-substrate
interface is located at $z=D$, with $-R<D<R$. The plane which passes
through the centre of the sphere is defined by $z=0$. The
dimensionless truncation parameter $t_r$
\begin{eqnarray}
  \label{truncation-parameter}
  t_r &=& \frac{D}{R}, \hspace{1cm} (-1<t_r<1),
\end{eqnarray}
describes the degree of truncation. For technical reasons which would
become apparent below, the part of the spheres lying below the
interface is introduced as a distinct medium~(medium~4). This latter
medium will finally be attributed properties identical to the
substrate~(medium~2). A main spherical coordinate system, ${\cal S}$,
is defined such that the origin of the radial coordinate $r$ coincides
with the centre of the sphere (Fig.~\ref{Fig:4}). The polar and
azimuthal angles are $\theta$ and $\phi$, respectively. The positive
$z$-axis, which is also the axis of revolution of the system, is
pointing downwards into the substrate. The impiging light of
wavelength $\lambda$ scatters at the surface at an incidence angle
$\theta_0$. To calculate the island polarizability, the polarization
of the islands by the incident light is modelled by using a multipolar
expansion and by introducing the image multipoles with respect to the
substrate~\cite{Wind87a}. The centre of the potential expansion is
chosen along the $z$-axis at a position $\mu R$ in the main coordinate
system ${\cal S}$, with 
\begin{eqnarray}
  \label{z-mp}
   -1< \mu < 1.
\end{eqnarray} 
The image multipole position becomes
$\bar{\mu} R$, where 
\begin{eqnarray}
  \label{z-imp}
  \bar{\mu} &=& 2 t_r - \mu.
\end{eqnarray}
It is convenient to introduce two other coordinate systems, ${\cal
    S}_\mu$ and ${\cal S}_{\bar{\mu}}$, whose origins are located at
the multipole and multipole image centres, respectively. The azimutal
angle of the plane of incidence of the light is defined as ${\phi_0}$.
It is taken equal to zero in all the numerical calculations which are
performed herein.

\subsection{Multipole expansions of the potential}
If the linear dimension of the island is small compared to the
wavelength of the incident light, retardation effects in the island
and around can safely be neglected. If the island material is assumed
to be non-magnetic, then the incident magnetic field will be
unaffected. Hence the main field is the electric field ${\bf E}({\bf
    r})$, which is related to the potential, $\psi({\bf r})$, in the
usual way ${\bf E}({\bf r}) = -{\mathbf \nabla} \psi({\bf r})$. When
retardation effects can be neglected or, in other words, when the size
of the island is supposed to be much smaller than the wavelength, this
potential must only satisfy the Laplace equation 
\begin{eqnarray} 
  \label{Laplacian-equation}
  {\mathbf \nabla}^2\psi({\bf r})  &=& 0.
\end{eqnarray} 
The appropriate boundary conditions for this potential come from the
continuity of the potential itself and of the normal components of the
displacement field 
\begin{mathletters}
\label{bound-cond}
\begin{eqnarray} 
  \psi_I({\bf r}_s)  &=& \psi_{II}({\bf r}_s), \\ 
  \varepsilon_I(\omega)\,\partial_n\psi_I({\bf r}_s)  &=& 
        \varepsilon_{II}(\omega)\,\partial_n\psi_{II}({\bf r}_s).
\end{eqnarray} 
\end{mathletters}
Here, $\varepsilon_i(\omega)$ is the frequency dependent dielectric
function of medium $i$, ${\bf r}_s$ is assumed to be any point on the
interface and $\partial_n$ denotes its normal derivative. These
conditions should be fulfilled for any two media having a common
interface.

To solve Eq.~(\ref{Laplacian-equation}), with the boundary conditions
Eqs.~(\ref{bound-cond}), it is convenient to use~\cite{Bedeaux88} a
multipole expansion for the potentials. In medium~1, the potential
then takes the form
\begin{eqnarray}
  \label{potential_1}
  \psi_1({\bf r}) &=& \psi_0({\bf r})
     + \sum_{lm}^{l\neq 0} A_{lm}\, r^{-l-1}_{\mu} Y_{lm}(\theta_{\mu}, \phi_{\mu})
     + \sum_{lm}^{l\neq 0} A^R_{lm} \, r^{-l-1}_{\bar{\mu}} 
            Y_{lm}(\theta_{\bar{\mu}}, \phi_{\bar{\mu}}).    
\end{eqnarray}
The spherical coordinates $(r_{\mu},\theta_{\mu},\phi_{\mu})$ and
$(r_{\bar{\mu}},\theta_{\bar{\mu}}, \phi_{\bar{\mu}})$ are referring
to the coordinate systems ${\cal S}_\mu$ and ${\cal S}_{\bar{\mu}}$,
which are centered at the points $\mu$ and $\bar{\mu}$, respectively
(Fig.~\ref{Fig:4}). The $m$-summation will be understood to range from
$m=-l$ to $l$. The spherical harmonics $Y_{lm}(\theta,\phi)$ are
normalized according to the convention~\cite{Morse} 
\begin{eqnarray}
  \label{spherical-harmonics}
  Y_{lm}(\theta,\phi) &=& \sqrt{ \frac{2\,l+1}{4\,\pi} \, \frac{(l-m)!}{(l+m)!} } \;
                         P^m_l(\cos\theta) 
                         \left( -1 \right)^m e^{im\phi},
\end{eqnarray}
where $P^m_l(\cos\theta)$ are the associated Legendre polynomials
defined as~\cite{Morse} 
\begin{eqnarray}
  \label{ass_legendre_A}
  P^m_l(x) &=& \frac{ (1-x^2)^{\frac{m}{2}} }{ 2^l\,l! }\; \frac{d^{\,l+m}}{dx^{\,l+m}} 
                \left(x^2-1\right)^l,
   \hspace{1cm}   x=\cos\theta,
\end{eqnarray}
for $m\geq 0$, while for $m<0$ 
\begin{eqnarray}
  \label{ass_legendre_B}
  P^m_l(x) &=&   \left(-1\right)^m \frac{(l+m)!}{(l-m)!}\;  P^{-m}_l(x).
\end{eqnarray} 
In Eq.~(\ref{potential_1}), $\psi_0({\bf r})$ stands for the potential
corresponding to the incident field ${\bf E}_0$ which, provided
clusters are small compared to the wavelength of the light, can be
approximated by a homogeneous field
\begin{eqnarray} 
  \label{initial-field}
  {\bf E}_0 = E_0
  (\sin\theta_0\cos\phi_0, \sin\theta_0\sin\phi_0, \cos\theta_0).
\end{eqnarray}
The potential $\psi_0({\bf r})$ related to this incident field,
then takes the following form 
\begin{eqnarray} 
  \label{initial-potential}
  \psi_0({\bf r}) 
         &=&  - r E_0\left(   \cos\theta \cos\theta_0 
                      + \sin\theta\cos\phi\sin\theta_0\cos\phi_0
                      + \sin\theta\sin\phi\sin\theta_0\sin\phi_0
               \right) \nonumber \\
         &=&  -r E_0 \sqrt{\frac{2\pi}{3}}
              \left[ \sqrt{2} \cos \theta_0 Y_{10}(\theta, \phi)
                    - \sin\theta_0 \left\{  e^{-i\phi_0} Y_{11}(\theta,\phi)   
                                          - e^{i\phi_0} Y_{1-1}(\theta,\phi) 
                                   \right\} 
              \right],
\end{eqnarray}   
with $E_0=|{\bf E}_0|$. The spherical coordinates
($r$,$\theta$,$\phi$) are defined with respect to the coordinate
system ${\cal S}$ whose origin is at the centre of the sphere.

For the potential inside the substrate, the solution of the Laplace
equation is chosen as 
\begin{eqnarray}
  \label{potential_2}
  \psi_2({\bf r}) 
     &=&  a_0 -r E_0 \sqrt{\frac{2\pi}{3}}
              \left[ \sqrt{2}\, a_1 \cos \theta_0 Y_{10}(\theta, \phi)
                    - \sin\theta_0 \left\{  a_2\, e^{-i\phi_0} Y_{11}(\theta,\phi)   
                                          - a_3\, e^{i\phi_0} Y_{1-1}(\theta,\phi) 
                                   \right\} 
              \right] \nonumber \\
     & & \mbox{} 
        + \sum_{lm}^{l\neq 0} A^T_{lm}\, r^{-l-1}_{\mu} Y_{lm}(\theta_{\mu}, \phi_{\mu}) ,
\end{eqnarray}
where $a_i$ are constants to be determined. Notice that the terms
inside the square brackets are a linear combination of the terms
involved in the expression~(\ref{initial-field}) of the potential
$\psi_0({\bf r})$. For the potential inside the cluster, medium~3 and
4, the following expansions are defined
\begin{mathletters}
  \label{potential_34}
  \begin{eqnarray}
  \label{potential_3}
  \psi_3({\bf r}) &=& b_0
  + \sum_{lm}^{l\neq 0} B_{lm}\, r^{l}_{\mu} Y_{lm}(\theta_{\mu}, \phi_{\mu})
     + \sum_{lm}^{l\neq 0} B^R_{lm} \, r^{l}_{\bar{\mu}} 
     Y_{lm}(\theta_{\bar{\mu}}, \phi_{\bar{\mu}}) \\
            \label{potential_4}
  \psi_4({\bf r}) &=& b'_0
  + \sum_{lm}^{l\neq 0} B^T_{lm}\, r^{l}_{\mu} Y_{lm}(\theta_{\mu},
     \phi_{\mu})
   \end{eqnarray}
\end{mathletters}
where $b_0$ and $b'_0$ are constants to be determined.

\subsection{Determination of the expansion coefficients via boundary conditions}
The various unknown expansion coefficients, $a_i$, $A_{lm}$, $B_{lm}$,
etc\ldots are now calculated in terms of the known parameters of the
model by imposing the boundary conditions on the considered geometry.
Since a formulation with image multipoles is used, the boundary
conditions at $z=D$ are easily satisfied. Their fulfilment leads to
the following relations
\begin{mathletters} 
\label{bc1}
\begin{eqnarray}
  \label{bc1a}
  a_0 &=& E_0\, d \left(\frac{\varepsilon_1}{\varepsilon_2}-1\right) \cos\theta_0,\\
  a_1 &=& \frac{\varepsilon_1}{\varepsilon_2}, \\
  a_2 &=& a_3 \;=\; 1, \\
  A^R_{lm} &=& \left(-1\right)^{l+m} 
             \frac{\varepsilon_1-\varepsilon_2}{\varepsilon_1+\varepsilon_2} A_{lm}, \\
  A^T_{lm} &=&  \frac{2\,\varepsilon_1}{\varepsilon_1+\varepsilon_2} A_{lm}. 
\end{eqnarray}
\end{mathletters} 
The continuity at the same boundary ($z=D$), but now inside the
sphere, gives 
\begin{mathletters} 
  \label{bc2}
\begin{eqnarray}
  \label{bc1d}
  b_0 &=&  b'_0 \\
  B^R_{lm} &=& \left(-1\right)^{l+m} 
             \frac{\varepsilon_3-\varepsilon_4}{\varepsilon_3+\varepsilon_4} B_{lm}, \\
  \label{bc1e}
  B^T_{lm} &=&  \frac{2\,\varepsilon_3}{\varepsilon_3+\varepsilon_4} B_{lm}. 
\end{eqnarray}
\end{mathletters} 
An explicit expression for the constant $b_0$ is given in
Eq.~(\ref{constant-term-above}) of Appendix~\ref{App:A}. These results
are obtained by taking advantage of both the orthogonality of the
spherical harmonics and the fact that on the surface of the substrate
($z=D$), as a consequence of the symmetry of the location of the
multipoles and image multipoles, the following relations hold
$r_\mu=r_{\bar{\mu}}$, $\theta_\mu = \pi - \theta_{\bar{\mu}}$ and
$\phi_\mu = \phi_{\bar{\mu}}$. Note that the
relations~(\ref{bc1})--(\ref{bc2}) are independant of the location of
the multipoles, so that they are similar to those obtained by Wind
{\it et al}.~\cite{Wind87a} under the assumption that the multipoles
are located at the centre of the sphere. Finally, there are only two
independant classes of expansion coefficients, namely $A_{lm}$ and
$B_{lm}$.

These coefficients can be derived from the relations that express the
fulfilment the boundary conditions at the surface of the sphere, where
$r=R$. By multiplying all the terms of those relations by the complex
conjugate spherical harmonic, $[Y_{lm}(\theta,\phi)]^*$, and by
integrating the resulting expressions over the surface of the sphere,
where one again takes advantage the orthogonality of the spherical
harmonics, the following (infinite) set of equations is obtained
\begin{mathletters} 
  \label{lin-system}
\begin{eqnarray}
  \sum^\infty_{l'=1} \left[   C^m_{ll'} R^{-l'-2} A_{l'm}
                           +  D^m_{ll'} R^{l'-1} B_{l'm} 
                     \right]   
         &=&  H^m_l ,  \\
  \sum^\infty_{l'=1} \left[   F^m_{ll'} R^{-l'-2} A_{l'm}
                           +  G^m_{ll'} R^{l'-1} B_{l'm} 
                     \right]   
         &=&  J^m_l, 
\end{eqnarray}
\end{mathletters} 
where $l=1,2,\ldots$, and $m=0,1$. The various matrix elements of the
above linear system can all be found in appendix~\ref{App:A}, where
their detailed derivation also is presented. 
The above linear system has only a non-trivial solution when $m=0$ and
$m={\pm 1}$~(cf. Eqs.~(\ref{App:right-Matrix-element-above}).  These
values are associated with the components of the uniform incident
field ${\bf E}_0$ which are orthogonal and parallel to the substrate,
respectively. An additional simplification can be made by showing that
the equations corresponding to $m=-1$ and $m=1$, respectively, are
equivalent.

To allow for a numerical solution of the infinite linear
system~(\ref{lin-system}), an upper cut-off $M$ in $l$ and $l'$ has to
be introduced. The integrals that define the system of
equations~(\ref{integrals}) are computed by Gauss-Legendre numerical
integration. Solutions of the linear system lead to values of the
expansion coefficients $A_{lm}$ and $B_{lm}$, which in turn can be
directly related to the (dipole) polarizability of the island.  Those
can be shown to be given by~\cite{Wind87b} 

\begin{mathletters}
  \label{pol}
\begin{eqnarray}
  \alpha_{\perp}  &=& \frac{2\pi\varepsilon_1 A_{10}}{
                        \sqrt{\frac{\pi}{3}} E_0 \cos\theta_0},  \\
  \alpha_{\parallel}  &=& -\frac{4\pi\varepsilon_1 A_{11}}{
                        \sqrt{\frac{2\pi}{3}} E_0 \sin\theta_0\exp(-i\phi_0)}.
\end{eqnarray}
\end{mathletters} 
Here, $\alpha_{\perp}$ and $\alpha_{\parallel}$ are the dipole
polarizabilities perpendicular and parallel to the interface of the
substrate, respectively. Due to the presence of the substrate, these
are in general quite different. In the present work, the parameter of
interest is the differential optical reflectivity $\Delta
R(\omega)/R(\omega)$ defined in Sec.~\ref{Sec:DifRef}. This quantity
is evaluated by modified Fresnel formulae for reflection where the
above given polarizabilities appear through the surface
susceptibilities~\cite{Bedeaux73}.

The present section was aimed at introducing a general method for the
calculation of the polarizability of truncated spheres supported by a
substrate. The model of Wind {\it et al}.~\cite{Wind87a,Wind87b} is
just a special case of this approach with $\mu=0$ (see
Appendix~\ref{App:A} for details), at least for $t_r \geq 0$. Indeed,
to obtain the solution for the case in which the centre of the sphere
is located below the substrate~($t_r<0$), these
authors~\cite{Wind87a,Wind87b} apply to the~($t_r>0$) case a
coordinate transformation method with a permutation of the dielectric
constants. This implies that different potential expansions are used
in the two cases, while, in the present work, the expansions in use do
not depend on the location of the multipoles with respect to the
surface.  Criteria are now needed to know how reliable are these
models, special attention being paid to cases corresponding to
negative values of the truncation ratio $t_r$.


\section{Fulfilment of the boundary conditions}
\label{Sec:NumRes}

In Sec.~\ref{Sec:DifRef} and in the already published
work~\cite{Remi}, the method has been only judged on the basis of the
convergence of the differential reflectivity curves $\Delta
R(\omega)/R(\omega)$. Indeed, in the model used herein, the
reflectivity only relies on the lowest order expansion coefficient
$A_{10}$ and $A_{11}$ (cf. Eqs.~(\ref{pol})). Therefore, the
occurrence of a convergence of the reflectivity curves upon, say,
increasing the multipole order $M$, does not provide any proof of the
reliability of the potentials since those depend on all the expansion
coefficients up to the cut-off order. The potentials given by the
expansions (\ref{potential_1}), (\ref{potential_2}) and
(\ref{potential_34}), can in principle be calculated in any point of
space provided the multipole coefficients, $A_{lm}$ and $B_{lm}$, are
known. Since the multipole expansions are indeed solutions of
Eqs.~(\ref{Laplacian-equation}) and (\ref{bound-cond}) at any $M$, it
is only the fulfilment of the boundaries conditions at the interface
between the various media that has to be checked. It has to be
stressed that this fulfilment is a more severe test for the
calculation than just considering the convergence of the differential
reflectivity curves. It is not only the self-consistency of the
numerical implementation but that of the method itself that is
controlled. Such a study thus provides a powerful and rigorous tool
for justifying the quality of previous and present numerical
simulations based on a multipole expansion.

The calculation including the multipoles and their images, the
boundary conditions at the cluster-substrate interface are obeyed by
construction. Thus, the only boundary conditions to be checked are at
the surface of the sphere. To measure the error in these, two error
functions are defined 
\begin{mathletters}
  \label{error}
\begin{eqnarray}
  {\cal E}_{\psi}({\bf r}_s) &=& 
       \frac{\psi_{+}({\bf r}_s)-\psi_{-}({\bf r}_s)}{
         \max_{{\bf r}_s} \psi_{0}({\bf r}_s) }, \\
  {\cal E}_{\partial_n \psi}({\bf r}_s) &=& 
       \frac{ \varepsilon_{+}\partial_n\psi_{+}({\bf r}_s)
                  -\varepsilon_{-}\partial_n\psi_{-}({\bf r}_s)}{
                 \max_{{\bf r}_s} \left( \varepsilon\partial_n\psi_{0}({\bf r}_s) \right) }.      
\end{eqnarray}
\end{mathletters}
The subscripts $+(-)$ correspond to quantity just outside~(inside) the
surface of the sphere. The maximum value of the incidence potential,
$\psi_0$, at the surface of the sphere, and the corresponding quantity
for the normal derivative are used as normalisation factors. The
impact of the parameters of the model, location of the multipole
centre, photon energy and multipole order on the accuracy of the
numerical calculations is examined in the following by considering the
values taken by the error functions in various representations of the
above mentioned Ag/MgO deposit.

\subsection{Location of the expansion centre}
In the model put forward in the present paper, the multipole expansion
point can move along the symmetry $z$-axis, instead of being at the
centre of the sphere as in the Bedeaux-Vlieger model. In
Fig.~\ref{Fig:5}, an attempt is made to determine the optimal position
$\mu$, for a given truncation ratio and value of $M =20$ at an energy
of $E=4.5$~eV.  The mean spatial absolute error in the boundary
conditions $\left<|{\cal E}_{\psi}({\bf r}_s)|\right>$
and$\left<|{\cal E}_{\partial_n \psi}({\bf r}_s)|\right>$, which are
defined in Eq.~(\ref{error}), are shown in Fig.~\ref{Fig:5} as
function of the truncation ratio. It can be observed that the value of
$\mu$ which leads to the minimum error depends on the truncation
ratio. For $t_r \geq 0$, the Bedeaux-Vlieger limit, where $\mu=0$,
provides the minimum error whereas, for $t_r \leq 0$, an improvement
is obtained by moving the expansion point inside the physical domain
or, in other words, by using negative $\mu$ values. However, for negative
truncation ratios, the matrix system become dramatically more and more
ill-conditioned, a fact that may be due to intervention of an
increasing power of distance between the surface of the sphere and the
expansion points $\mu$ and $\bar{\mu}$.

In principle, the infinite expansion of potentials on which the method
is based should lead to a unique numerical value, wherever the
location of the expansion centre is. In the case of numerical
calculations, where the expansion is truncated at an order $M$, it is
not too surprising to find, for a given value of $M$, a (weak)
dependency of the error functions on the position of the expansion
point, as can be already stressed in Fig.~\ref{Fig:5}. A search for an
optimal position of the expansion centre can be performed for each
value of the truncation ratio (Fig.~\ref{Fig:6}). For $t_r \geq 0$
(see also Fig.~\ref{Fig:5}), it is the centre of the sphere ($\mu=0$).
For $t_r \leq 0$, it appears that the best choice is to place the
$\mu$ point close to the substrate so that $\mu=t_r$, that is to say
to superimpose the centre of expansion and its image. Roughly
speaking, for $t_r\leq0$, between one and two order of magnitude in
the error function can be earned by the present method with respect to
the Bedeaux-Vlieger model in which $\mu=0$.

\subsection{Boundary conditions at the resonance energy}
When the energy of the incident light is close to a resonance of the
system, i.e.\, in our case, close to either the low energy resonance around
$2.5$~eV or the high energy resonance at $3.7$~eV, the boundary
conditions are even harder to satisfy. This is seen from
Fig.~\ref{Fig:7} where the (spatial) mean of the absolute error in the
boundary conditions, $\left<|{\cal E}_{\psi}({\bf r}_s)|\right>$ and
$\left<|{\cal E}_{\partial_n \psi}({\bf r}_s)|\right>$, are shown as
function of energy $E=\hbar\omega$ for truncation ratio $t_r=0.1$ and
$t_r=-0.1$. The dips which can be observed around the low energy
resonance at $E=2.5$~eV are believed to come from numerical artefacts.
These figures highlight the sizable effect of the dielectric function
on the quality of the simulation, since the energy range $E=2$~eV to
$E=3.7$~eV, where the resonances are peaking, concentrates the highest
values of the error functions. This is, however, probably not so
surprising in view of the fact that the system is close to singular at
a resonance point. As the truncation ratio is reduced, the overall
error seems to be increased, but it is still the regions around the
resonances that are associated with the largest errors. Roughly
speaking, the errors in the boundary conditions are raised by an
energy independant factor as the truncation ratio is reduced
(Fig.~\ref{Fig:7}).

\subsection{Multipole order}
By increasing the number of multipoles included in the calculation,
$M$, the fulfilment of the boundary conditions should be improved
since the size of the function basis increases. Fig.~\ref{Fig:8} shows
the errors $\left<|{\cal E}_{\psi}({\bf r}_s)|\right>$ and
$\left<|{\cal E}_{\partial_n \psi}({\bf r}_s)|\right>$ as function of
$M$ for the energy $E=4.5$~eV and $\mu=0$ for two truncation ratios
$t_r=0.1$ and $t_r=-0.1$. These figures are representative of the
cases where $\mu=0$. For $t_r \geq 0$, the error decreases upon
increasing $M$, which corresponds to the expected behaviour. However,
for $t_r \leq 0$, the error is not seen to decrease upon increasing
$M$. In this case, the error functions are quite sensitive to the
value of $\mu$. This is illustrated in Fig.~\ref{Fig:9} where the
error functions are given for $\mu$-values of $0.$ and $-0.1$. It is
observed that moving the expansion point away from the center of the
sphere may dramatically improve the overall error made in the
simulation. Finally, it can be noted that, in all cases, a higher
limit in $M$ is imposed by the vanishing of the matrix conditionning.

\subsection{Spatial variations of the error functions}
The evolutions of the potential $\psi$ and of its normal derivative
$\varepsilon \partial_n\psi$ along the surface of the sphere, in the
plane of incidence of the light (Fig.~\ref{Fig:10}a and
Fig.~\ref{Fig:10}b) and in a plane perpendicular to it
(Fig.~\ref{Fig:10}c and Fig.~\ref{Fig:10}d), are given as function of
the polar angle $\theta$ in Fig.~\ref{Fig:10} for the case
corresponding to $t_r=0.1$, $E=4.5$~eV, $\mu=0$ and $M=16$. 
Also shown  are the error functions 
 $\left<|{\cal E}_{\psi}({\bf r}_s)|\right>$ 
(Fig.~\ref{Fig:10}a and Fig.~\ref{Fig:10}c) and
$\left<|{\cal E}_{\partial_n \psi}({\bf r}_s)|\right>$
(Fig.~\ref{Fig:10}b and Fig.~\ref{Fig:10}d) in the bottom panels of
the graphs. For the same set of parameters, we have also found
(results not shown) that the errors are slowly varying function of the
azimuthal angle $\phi$.  The regions which give rise to the largest
errors in the boundary conditions are concentrated around the top of
the sphere and, to some extent, at the interface between the spherical
cap and the substrate.  The same trends can be observed in the global
view of the boundary conditions which is presented in
Fig.~\ref{Fig:11} for the same case. This view consists in a
projection in the plane of incidence of 
$\left<|{\cal E}_{\psi}({\bf r}_s)|\right>$ 
(upper panel) and
$\left<|{\cal E}_{\partial_n \psi}({\bf r}_s)|\right>$
The direction of the incidence of the light is indicated by an arrow.
The error is roughly of the same order of magnitude all over the
sphere.


\section{Conclusions}
A generalised method for the calculation of the polarizability of a
truncated sphere by means of a multipolar description has been
presented. The proposed approach, which is an extension of earlier
published works due to Bedeaux, Vlieger and
co-workers~\cite{Wind87a,Wind87b}, and which contains this model as a
special case, is based on the possibility for the multipole expansion
centre to freely move along the $z$-axis of the main coordinate system
instead of being located at the centre of the sphere. This method
allows to perform the calculations for cases in which the numerical
code is badly conditioned when the centre of the truncated sphere is
below the plane of truncation. Let us stress that this geometry has an
strong importance since it corresponds to clusters whose aspect ratio
is higher than 2.

Neither the convergence of the model itself nor the agreement between
the data and the model can guarantee that the solution is correct,
since the reflectivity curves only depend on the lowest order
expansion coefficients. A way to assess the quality of the numerical
calculation up to the upper order has been introduced which consists
in judging the accuracy of the models on the basis of the fulfilment
of the appropriate boundary conditions. It is shown that, provided
the cluster aspect ratio is not too high, the multipole expansion
method represents a rather accurate tool for the determination of mean
values of the parameters characterizing the deposited clusters. In all
cases however, the values found for the error functions allow to
estimate the validity of the model. The simulation of the optical
reflectivity in the UV-visible range allows to determine cluster size,
shape ratio and density in a rather accurate manner, as illustrated
herein by the case of a Ag/MgO(100) deposit. The method offers a
promising tool for the {\it in situ} examination of cluster growth on
substrates.

\acknowledgements One of the authors (I.S.) would like to thank
D.~Bedeaux for invaluable discussions and for kindly making available
unpublished notes regarding the topic of thin discontinuous films.
 
I.S. would like to thank the Research Council of
Norway~(Contract No.  32690/213), Norsk Hydro ASA, Total Norge ASA,
CNRS~(Centre National de la Recherche Scientifique) and R.L. would aknowledge Saint-Gobain
Recherche for financial support.


\appendix

\section{The matrix elements}
\label{App:A}

In this appendix we give the various matrix elements, and the form of
the right hand side, appearing in the matrix
system~(Eqs.~\ref{lin-system}). With the particular expansions chosen
for the potentials $\psi_i({\bf r})$ ($i=1,2,3,4)$ (cf.\ 
Sec.~\ref{Sec:Model}), and the relations~Eqs.~(\ref{bc1}) and
(\ref{bc2}), the boundary conditions at the surface of the substrate,
by construction, are automatically fulfilled.  However, we still need
to satisfy the remaining boundary conditions~Eqs.(\ref{bound-cond}) on
the surface of the sphere. By taking these boundary conditions,
multiplying by the complex conjugate spherical harmonic,
$[Y_{lm}(\theta,\phi)]^*$, and integrating over all directions, i.e.\ 
using the weak formulation of the boundary conditions, one is lead to
the following matrix system (cf.\ Eqs.~(\ref{lin-system})) 
\begin{mathletters} 
  \label{App:lin-system}
\begin{eqnarray}
  \sum^\infty_{l'=1} \left[   C^m_{ll'} R^{-l'-2} A_{l'm}
                           +  D^m_{ll'} R^{l'-1} B_{l'm} 
                     \right]   
         &=&  H^m_l ,  \\
  \sum^\infty_{l'=1} \left[   F^m_{ll'} R^{-l'-2} A_{l'm}
                           +  G^m_{ll'} R^{l'-1} B_{l'm} 
                     \right]   
         &=&  J^m_l.
\end{eqnarray}
\end{mathletters} 
This linear system can be used to determine the expansion coefficients
$A_{lm}$ and $B_{lm}$ for all allowed values of $l$ and $m$; $l=0,\pm
1,\pm 2, \ldots$ and $m=0,\pm1,\pm2,\ldots,\pm m$.  However, for
orthogonality reasons on integration over $\phi$, this system is
reduced to $m=-1,0,1$, since only the (uniform) incident field
contains these ``quantum numbers''~(cf. Eq.~(\ref{initial-potential}).  The
matrix-elements of the system~(Eqs.~\ref{App:lin-system}) are thus given by
the  following expressions 
\begin{mathletters} 
  \label{Matrix-element-above}
\begin{eqnarray}
  \label{C-Matrix-element}
  C^{m}_{ll'}  &=&  \zeta^{m}_{ll'} 
     \left[ 
          K^{m}_{ll'}[\mu](t_r) 
        + \frac{\varepsilon_1-\varepsilon_2}{\varepsilon_1+\varepsilon_2}
           \left(-1\right)^{l'+m} \;
                                     K^{m}_{ll'} [\bar{\mu}](t_r)
        + \frac{2\,\varepsilon_1}{\varepsilon_1+\varepsilon_2} 
             \left\{ K^{m}_{ll'}[\mu](t_r=1)-K^{m}_{ll'}[\mu](t_r) \right\}
     \right] ,  \\
  \label{D-Matrix-element}
  D^{m}_{ll'}  &=&  -\zeta^{m}_{ll'} 
     \left[ 
          M^{m}_{ll'}[\mu](t_r) 
        + \frac{\varepsilon_3-\varepsilon_4}{\varepsilon_3+\varepsilon_4}\left(-1\right)^{l'+m} \;
                                     M^{m}_{ll'} [\bar{\mu}](t_r)
        + \frac{2\,\varepsilon_3}{\varepsilon_3+\varepsilon_4} 
             \left\{ M^{m}_{ll'}[\mu](t_r=1)-M^{m}_{ll'}[\mu](t_r) \right\}
     \right] ,  \\
  \label{F-Matrix-element}
  F^{m}_{ll'}  &=&  \zeta^{m}_{ll'} 
     \left[ 
          \frac{2\,\varepsilon_1 \varepsilon_2}{\varepsilon_1 +\varepsilon_2}\; L^{m}_{ll'}[\mu](t_r=1)
        + \varepsilon_1\frac{\varepsilon_1 -\varepsilon_2}{\varepsilon_1 +\varepsilon_2}
          \left\{ L^{m}_{ll'}[\mu](t_r) + \left(-1\right)^{l'+m} L^{m}_{ll'} [\bar{\mu}](t_r)\right\}
     \right] ,
\end{eqnarray}
and  
\begin{eqnarray}
  \label{G-Matrix-element}
  G^{m}_{ll'}  &=& - \zeta^{m}_{ll'} 
     \left[ 
          \frac{2\,\varepsilon_3 \varepsilon_4}{\varepsilon_3 +\varepsilon_4} \;N^{m}_{ll'}[\mu](t_r=1)
        + \varepsilon_3\frac{\varepsilon_3 -\varepsilon_4}{\varepsilon_3 +\varepsilon_4}
          \left\{ N^{m}_{ll'}[\mu](t_r) + \left(-1\right)^{l'+m} N^{m}_{ll'} [\bar{\mu}](t_r)\right\}
     \right] .
\end{eqnarray} 
\end{mathletters} 
Here the following notation has been introduced 
\begin{eqnarray}
  \label{zeta}
  \zeta^{m}_{ll'}  &=&  \frac{1}{2}
     \sqrt{ \frac{ (2l+1)\,(2l'+1)\,(l-m)!\,(l'-m)!}{(l+m)!\,(l'+m)!} }  ,
\end{eqnarray}
and the integrals $K$, $L$, $M$, and $N$ will be defined below.
Furthermore, the right-hand side of the system
Eqs.~(\ref{App:lin-system}) is defined as 
\begin{mathletters} 
  \label{App:right-Matrix-element-above}
\begin{eqnarray}
  \label{H-Matrix-element-above}
  H^{m}_{l}  &=& 
      \sqrt{4\pi} 
      \left[ 
         \frac{b_0}{R} - E_0t_r\frac{\varepsilon_1-\varepsilon_2}{\varepsilon_2}\cos\theta_0
      \right] \delta_{0l} \delta_{0m} 
    \nonumber\\ \mbox{} & &    
    + \sqrt{\frac{4\pi}{3}} E_0\cos\theta_0 \, \delta_{0m}
      \left[ \frac{\varepsilon_1}{\varepsilon_2}\, \delta_{1l}
            +\frac{\varepsilon_1-\varepsilon_2}{\varepsilon_2}
             \left\{
                 \sqrt{3}t_r \zeta^0_{l0} Q^0_{l0}(t_r) - \zeta^0_{l1} Q^0_{l1}(t_r)
             \right\}
      \right]   
    \nonumber \\ \mbox{} & &    
    - \sqrt{\frac{2\pi}{3}} E_0\cos\theta_0 \, \delta_{l1}
      \left[ e^{-i\phi_0}\,\delta_{1m} -  e^{i\phi_0}\, \delta_{-1m}
      \right] ,
\end{eqnarray}
and 
\begin{eqnarray}
  \label{J-Matrix-element-above}
  J^{m}_{l}  &=& 
      \sqrt{\frac{4\pi}{3}} E_0 \varepsilon_1 \cos\theta_0\,  \delta_{0m} \delta_{1l}
    - \sqrt{\frac{2\pi}{3}} E_0 \varepsilon_2 \sin\theta_0\, \delta_{1l}
        \left[ e^{-i\phi_0}\,\delta_{1m} -  e^{i\phi_0}\, \delta_{-1m} \right]
  \nonumber \\ \mbox{} & & 
    - \sqrt{\frac{2\pi}{3}}\, \left[ (\varepsilon_1-\varepsilon_2) E_0\sin\theta_0 \right] 
       \left[  e^{-i\phi_0} \zeta^{1}_{1l} Q^1_{l1}(t_r)\, \delta_{1m}
             - e^{i\phi_0} \zeta^{-1}_{l1} Q^{-1}_{l1}(t_r)\, \delta_{-1m}
        \right] . 
\end{eqnarray}
\end{mathletters} 
The above equations depend on several types of integrals. They are defined as 
\begin{mathletters} 
  \label{integrals}
\begin{eqnarray} 
  \label{Q-int}
  Q^{m}_{ll'} (t_r) &=& \int^{t_r}_{-1}\!dx\; P^m_l(x) P^m_{l'}(x), \\
  \label{K-int}
  K^{m}_{ll'} [\eta] (t_r) &=& \int^{t_r}_{-1}\!dx\; 
                  P^m_l\left( x \right) 
                  P^m_{l'}\left( 
                             \frac{x-\eta(t_r)}{\sqrt{ \chi[\eta](x,1) }}
                           \right)
                  \left[ \chi[\eta](x,1) \right]^{\frac{-l'-1}{2}} , \\
  \label{L-int}
  L^{m}_{ll'} [\eta] (t_r) &=& \int^{t_r}_{-1}\!dx\; 
              \left. \left[    
                  P^m_l\left( x \right) 
                  \partial_r \left\{
                              P^m_{l'}\left( 
                                 \frac{xr-\eta(t_r)}{\sqrt{\chi[\eta](x,r}}
                              \right)
                             \left[ \chi[\eta](x,r)\right ]^{\frac{-l'-1}{2}}
                             \right\}
               \right] \right|_{r=1} , \\
  \label{M-int}
  M^{m}_{ll'} [\eta|] (t_r) &=& \int^{t_r}_{-1}\!dx\; 
                  P^m_l\left( x \right) 
                  P^m_{l'}\left( 
                             \frac{x-\eta(t_r)}{\sqrt{\chi[\eta](x,1)}}
                           \right)
                  \left[ \chi[\eta](x,1) \right]^{\frac{l'}{2}}, 
\end{eqnarray} 
and
\begin{eqnarray}
  \label{N-int}
  N^{m}_{ll'} [\eta] (t_r) &=& \int^{t_r}_{-1}\!dx\; 
              \left. \left[    
                  P^m_l\left( x \right) 
                  \partial_r \left\{
                              P^m_{l'}\left( 
                                 \frac{xr-\eta(t_r)}{\sqrt{ \chi[\eta](x,r)}}
                              \right)
                             \left[ \chi[\eta](x,r)
                             \right]^{\frac{l'}{2}} \right\} \right]
                             \right|_{r=1}.
\end{eqnarray} 
\end{mathletters} 
All the above integrals, except the Q-integral, have a functional
dependency on the function $\eta(t_r)$, which is used as a generic
case for either $\mu(t_r)$ or $\bar{\mu}(t_r)$, i.e.\ for the
$z$-coordinate of the position of the multipoles or image multipoles
in the main coordinate system ${\cal S}$ with the origin in the center
of the sphere.  Moreover, the functional, $\chi[\eta](x,r)$, appearing
in Eqs.~(\ref{K-int})--(\ref{N-int}), is defined as
\begin{eqnarray}
  \label{chi}
  \chi[\eta](x,r) &=& r^2-2\eta(t_r)xr+\eta^2(t_r), \hspace{1.1cm} x=\cos\theta.
\end{eqnarray}
This functional gives the distance to a given point from the location
of the multipoles~($\eta=\mu$) or image multipoles~($\eta=\bar{\mu}$)
in terms of the coordinates, $(r,\theta,\phi)$, of the main coordinate
system ${\cal S}$. Notice that $\chi[\eta](x,r)$ is independent of the
azimuthal angle $\phi$. This is, of course, a consequence of the
rotational symmetry around the z-axis of the considered geometry.

We would like to stress that with Eq.~(\ref{ass_legendre_B}) and the
definition of $\zeta_{ll'}^m$, Eq.~(\ref{zeta}), the matrix elements
constituting the left-hand-side of our linear
system~Eqs.~(\ref{Matrix-element-above}), do not depend on the actual
sign of $m$, but only its value, i.e.\  $C^m_{ll'}=C^{-m}_{ll'}$ with
similar expression for the other (left hand-side) matrix elements. For
the right-hand-side matrix elements, it follows directly from the
definitions of these quantities,
Eqs.~(\ref{App:right-Matrix-element-above}), that they vanish
identically for $m\neq 0,\pm 1$ and that they furthermore satisfy the
relations $H^1_l\exp(i\phi_0)=-H^{-1}_l\exp(i\phi_0)$ and
$J^1_l\exp(i\phi_0)=-J^{-1}_l\exp(i\phi_0)$. This is so because the
right-hand-side of the linear system stands entirely for the incident
field, which is the same in the present case and the one considered in
Ref.~\cite{Wind87a}, and therefore is independant of the location of
the multipoles. As a consequence it follows that the multipole
coefficients fullfil the following relations 
\begin{eqnarray}
  A_{l1}\, e^{i\phi_0} &=& - A_{l-1}\, e^{-i\phi_0}, \\
  B_{l1}\, e^{i\phi_0} &=& - B_{l-1}\, e^{-i\phi_0}. 
\end{eqnarray}
Hence, the independent multipole coefficients are chosen to those
corresponding to $m=0,1$. All other coefficients either vanish or can
be expressed in terms of these. Furthermore, it can be shown that if
the physical system does not contain any free charges, then the
equations with $l=0$ is useless. Thus the linear
system~Eqs.~(\ref{App:lin-system}) will only be of interest when
$l=1,2,3,\ldots$ and $m=0,1$.  The above findings look like those
found in the Bedeaux-Vlieger model~\cite{Wind87a}.

We may take the ``Bedeaux-Vlieger'' limit of our results by placing
the multipoles in the center of the sphere, i.e.\ 
\begin{eqnarray}
  \label{WVB-limit}
  \mu(t_r) &=& 0, \\
   \bar{\mu}(t_r) &=& 2t_r.
\end{eqnarray}
Using the notation of Wind {\it et al}.~\cite{Wind87a,Wind87b}, the
above introduced integrals then take the following form
\begin{mathletters}
  \label{WVB-limit-int}
\begin{eqnarray}
  K^m_{ll'}[\mu](t_r)       =   Q^m_{ll'}(t_r),   & \hspace{2cm} &
         M^m_{ll'}[\mu](t_r)       =   Q^m_{ll'}(t_r), \\
  K^m_{ll'}[\bar{\mu}](t_r) =   S^m_{ll'}(t_r) ,  & \hspace{2cm} &
        M^m_{ll'}[\bar{\mu}](t_r)  =   T^m_{ll'}(t_r), \\
  L^m_{ll'}[\mu](t_r)       =   -(l'+1) \, Q^m_{ll'}(t_r), & \hspace{2cm} &
        N^m_{ll'}[\mu](t_r)        =   l' Q^m_{ll'}(t_r), \\
  L^m_{ll'}[\bar{\mu}](t_r) =    \partial_t \left.S^m_{ll'}(t,t_r)\right|_{t=1}, & \hspace{2cm} &
        N^m_{ll'}[\bar{\mu}](t_r)  =   \partial_t \left. T^m_{ll'}(t_r) \right|_{t=1}. 
\end{eqnarray}
\end{mathletters}
The $Q^m_{ll'}$-integral, which is independent of the position of the
multipoles, is the same in both works. The linear system, of
Ref.~\cite{Wind87a}, used to determine the multipole expansion
coefficients, can now be obtained from our
formulae~Eqs.~(\ref{App:lin-system}) by taking advantage of the
relations~Eqs.~(\ref{WVB-limit-int}). In this limit our formulae for
the linear system reduce, as they should, to those presented by Wind
{\it et al}.~\cite{Wind87a,Wind87b}.

In order to evaluate the potentials in the check of the boundary
conditions, the constant $b_0$, appearing in Eq.~(\ref{potential_3}),
must be determined. This constant is obtained in the same way as the
above matrix elements, but here only terms containing
$Y^*_{0,0}(\theta,\phi)$ will contribute. The result is 
\begin{eqnarray}
  \label{constant-term-above}
  b_0 &=&  \sqrt{\frac{R}{3}} E_0 \left( \frac{\varepsilon_1}{\varepsilon_2}-1\right)   
                  \cos\theta_0 \zeta^0_{01} Q^0_{01}(t_r)
           + R E_0 t_r \left( \frac{\varepsilon_1}{\varepsilon_2}-1\right)
                  \cos\theta_0\left[1-\zeta^0_{00}Q^0_{00}(t_r)   \right]   
         \nonumber \\ && \mbox{}
           + \frac{1}{\sqrt{4\pi}} \sum^{\infty}_{l'=1} A_{l'0} R^{-l'-1}\zeta^0_{0l'}
                 \left[ K^0_{0l'}[\mu](t_r) + (-1)^{l'}
                       \frac{\varepsilon_1-\varepsilon_2}{\varepsilon_1+\varepsilon_2}
                         K^0_{0l'}[\bar{\mu}](t_r) 
                      + \frac{2\varepsilon_1}{\varepsilon_1+\varepsilon_2}
                         \left\{ K^0_{0l'}[\mu](1) - K^0_{0l'}[\mu](t_r) \right\} \right]
         \nonumber \\ && \mbox{}
           -\frac{1}{\sqrt{4\pi}} \sum^{\infty}_{l'=1} B_{l'0} R^{l'}\zeta^0_{0l'}
                 \left[ M^0_{0l'}[\mu](t_r) + (-1)^{l'}
                       \frac{\varepsilon_3-\varepsilon_4}{\varepsilon_3+\varepsilon_4}
                         M^0_{0l'}[\bar{\mu}](t_r) 
                      + \frac{2\varepsilon_3}{\varepsilon_3+\varepsilon_4}
                         \left\{ M^0_{0l'}[\mu](1) - M^0_{0l'}[\mu](t_r) \right\} \right].
\end{eqnarray}



\widetext
\newpage
\begin{figure}[h]
    \caption{Experimental data for a 2 nm thick silver deposit on a magnesium 
        oxide~MgO(100) substrate: (a) A FEG-SEM image of the film.
        The horizontal bar indicated in the figure corresponds to
        200~nm; (b) The experimental differential reflectivity
        spectrum obtained at an incident angle of $45^\circ$ (circular
        symbols). A simulation result, using the method of
        Ref.~\protect\cite{Wind87a,Wind87b}, is indicated by a solid
        line, and demonstrate good agreement with the experimental
        data. In the simulation, the truncated spherical particles, of
        radius $R=6.8$~nm, were placed on a regular grid of lattice
        constant $19.6$~nm, and the truncation parameter (see text)
        was $t_r=0.11$~(corresponding to a aspect ratio of $1.80$).
        Notice that the main spectral features and the numerical
        values of both the low and high-energy resonances are well
        predicted by the theoretical model.}
    \label{Fig:1}
\end{figure}

\begin{figure}[h]
    \caption{Simulated differential reflectivity curves 
        $\Delta R(\omega)/R(\omega)$ as a function of energy
        $E=\hbar\omega$ for differents number of multipoles included
        in the simulation. Clusters are defined by the same numerical
        parameters as in Fig.~\protect\ref{Fig:1}.}
    \label{Fig:2}
\end{figure}

\begin{figure}[h]
    \caption{Simulated differential reflectivity by the Bedeaux-Vlieger
        truncated spherical model for the same islands parameters as
        in Fig.~\protect\ref{Fig:1} but for a negative truncation
        ratio $t_r=-0.11$~(aspect ratio of $2.25$).  The various
        curves correspond to different choices for the multipole order
        as indicated in the legend of the figure. Notice that in the
        region around the low energy resonance no convergence seems to
        be reached by increasing the number of multipoles included in
        the calculation.  The inset shows the details of the
        differential reflectivity curves around the low energy
        resonance.}
    \label{Fig:3}
\end{figure}

\begin{figure}[h]
    \caption{The cross section of the geometry considered in the 
      present work. Here $\mu R$ indicates the z-coordinate of the
      location of the multipoles, while $\bar{\mu} R$ is the same
      quantity, but for the image multipoles. Notice that the
      multipoles and thus also the image multipoles, are always
      located on the z-axis which is the axis of revolution. The
      position of the substrate is parallel to the xy-plane, and it
      is located at $z=D$, where $D$ is a (signed) real constant. The
      dielectric functions of the various regions are those indicated
      in the figure by $\varepsilon_i(\omega)$, with $i=1,\ldots,4$.}
    \label{Fig:4}
\end{figure}

\begin{figure}[h]
    \caption{The mean errors as defind in the text, 
        $\left<|{\cal E}_{\psi}({\bf r}_s)|\right>$ and $\left<|{\cal
              E}_{\partial_n \psi}({\bf r}_s)|\right>$, in the
        boundary conditions of (a) the potential and (b) its normal
        derivative, as function of $t_r$ for different $\mu$ values.
        The model clusters are defined with the same parameters as in
        the Ag/MgO(100) case (Fig.~\ref{Fig:1}), with $E=4.5$~eV and
        $M=20$.}
    \label{Fig:5}
\end{figure}

\begin{figure}[h]
    \caption{Fulfilment of the boundary conditions as function of
        $\mu$ for an energy $E=4.5$~eV, a multipole order $M=20$ and
        two given values of $t_r$, $0.1$ and $-0.1$. Error functions
        on (a) the potential and (b) the normal derivative.}
    \label{Fig:6}
\end{figure}

\begin{figure}[h]
    \caption{The mean error in the boundary conditions as a function 
        of energy for two values of $t_r=0.1$ and $t_r=-0.1$, in the
        case defined by $\mu=0$ and $M=20$, for (a) the potential and
        (b) the normal derivative.}
    \label{Fig:7}
\end{figure}

\begin{figure}[h]
    \caption{The evolution of the error in the boundary conditions as 
        a function of the order in the multipole development in the
        case of which $E=4.5$ eV, $\mu=0$, for two given values of
        $t_r$, $0.1$ and $-0.1$ ((a) potential (b) the normal
        derivative).}
    \label{Fig:8}
\end{figure}

\begin{figure}[h]
    \caption{The improvment brought in the convergence of the
        boundary conditions by moving the expansion point in the
        case $t_r=-0.1$, $E=4.5$~eV as a function of the multipole
        order $M$.}
    \label{Fig:9}
\end{figure}

\begin{figure}[h]
  \caption{Evolution of the potential (a and b) and its normal
      derivative (c and d) along the surface of the sphere. The errors
      in the boundary conditions of the potential on the surface of
      the sphere, ${\cal E}_{\psi}({\bf r}_s)$~(a and c) and of the
      normal derivative ${\cal E}_{\partial_n \psi}({\bf r}_s)$~(b and
      d), are shown as function of the spherical coordinate $\theta$
      for given value of the azimutal angle $\phi$. The values used
      for the angle $\phi$ correspond to the incident
      plane~($\phi=0$)~(a and b) and a plane perpendicular to the
      incident plane~($\phi=\pi/2$)~(c and d). The energy of the
      incident light is $E=4.5$~eV, the truncation ratio is $t_r=0.1$
      (and $\mu=0.$) and the number of multipoles is $M=16$.}
    \label{Fig:10}
\end{figure}

\begin{figure}[h]
    \caption{Global view of the boundary condition errors 
         (a)~$\left<|{\cal E}_{\psi}({\bf r}_s)|\right>$ and 
         (b)~$\left<|{\cal E}_{\partial_n \psi}({\bf r}_s)|\right>$ 
         when projected into the plane of incidence.
         The parameters used are $E=4.5$~eV, $t_r=0.1$, and  $M=20$.}
    \label{Fig:11}
\end{figure}

\newpage

\setcounter{figure}{1}
\newcommand{\mycaption}[2]{\begin{center}{\bf Figure \thefigure}\\{#1}\\{\em #2}\end{center}\addtocounter{figure}{1}}
\newcommand{\myauthor}{I.\ Simonsen, R.\ Lazzari, J.\ Jupille and S.\ Roux}
\newcommand{\mytitle}{Optical response of supported particles}

\begin{figure}[h]
  \begin{center}
    \Large a) \epsfig{file=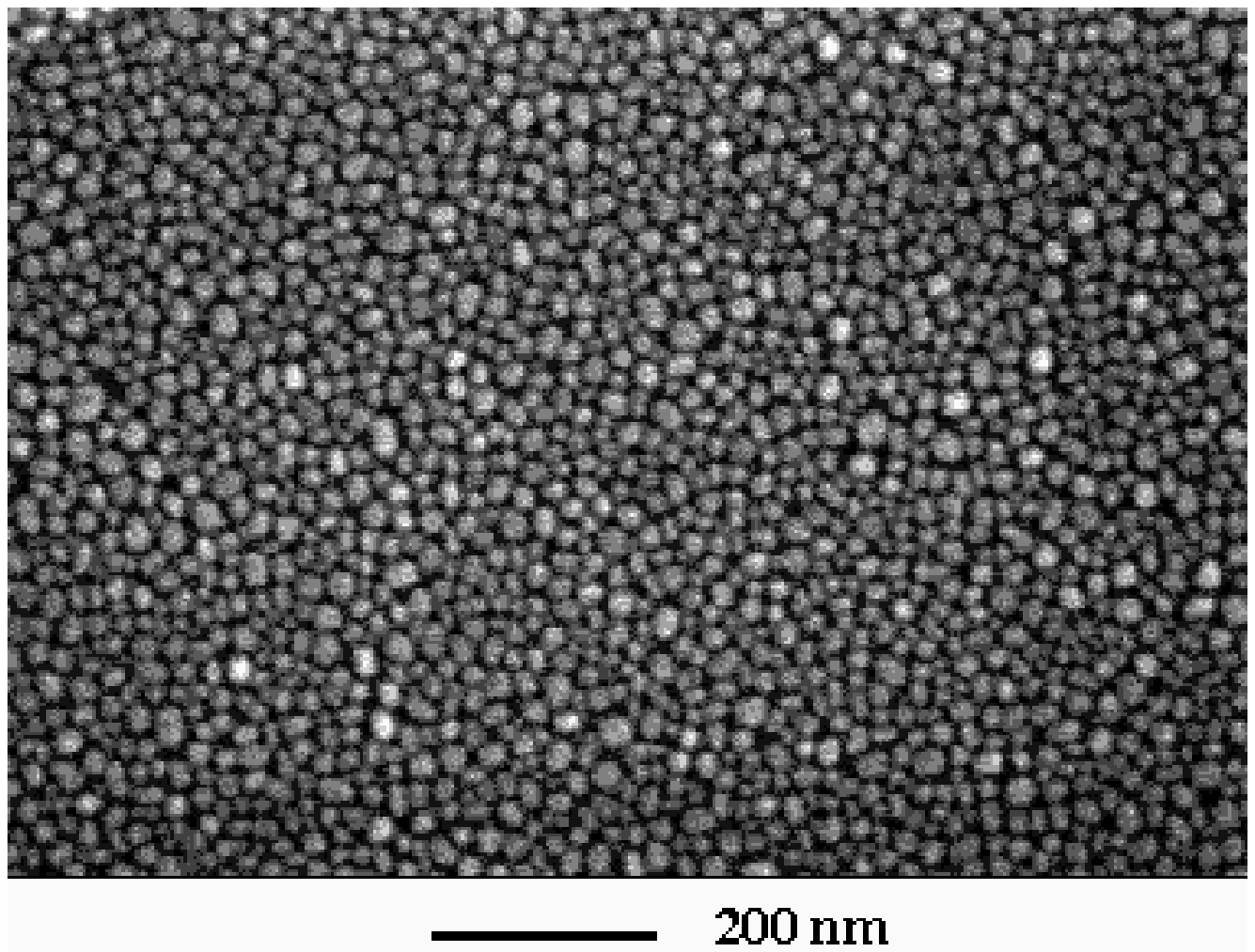,width=12cm} \vspace*{0.7cm} \\ 
    \Large b) \epsfig{file=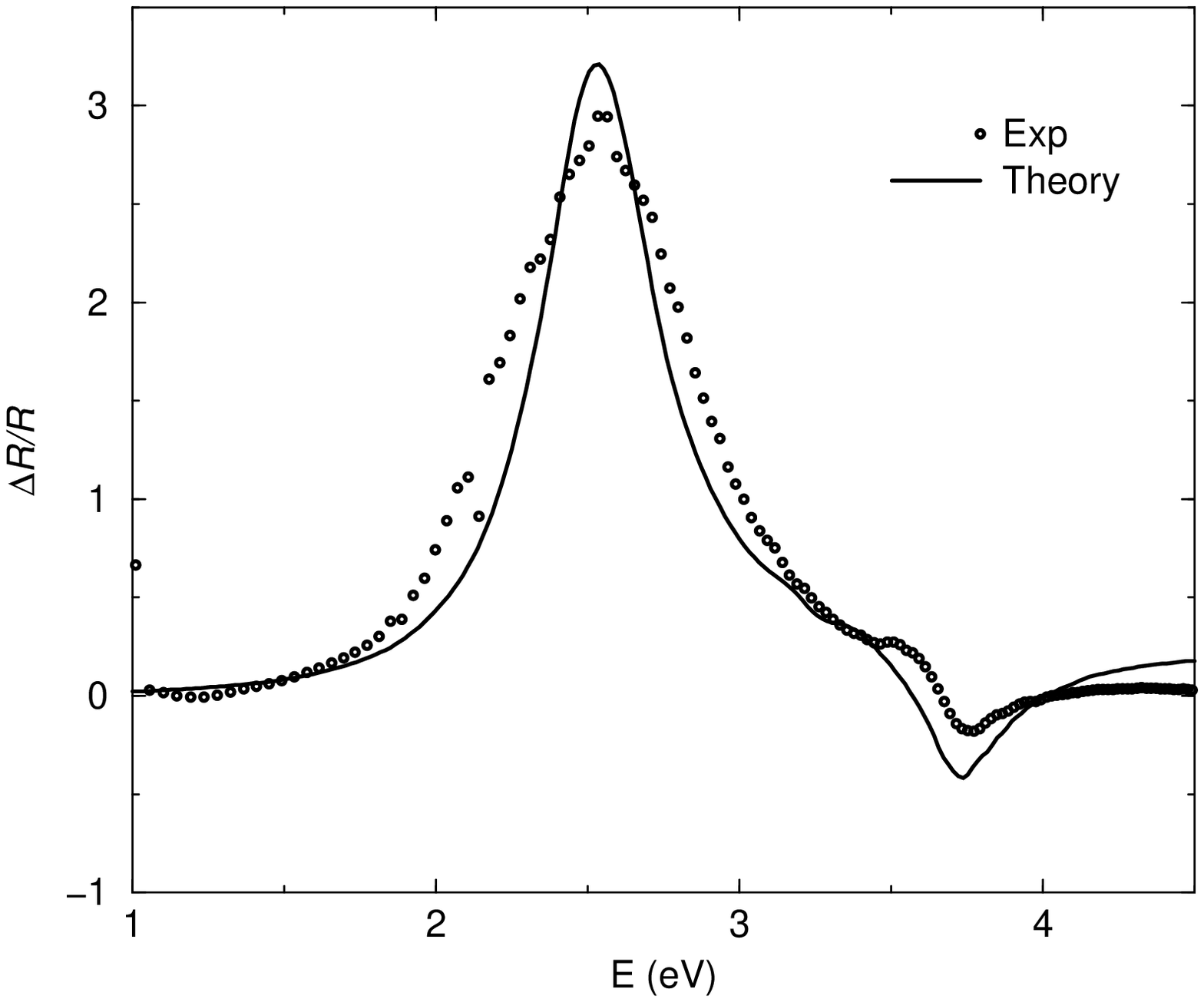,width=12cm} 
  \end{center} 
    \mycaption{\myauthor}{\mytitle}
\end{figure}

\newpage

\begin{figure}[h]
  \begin{center}
    \epsfig{file=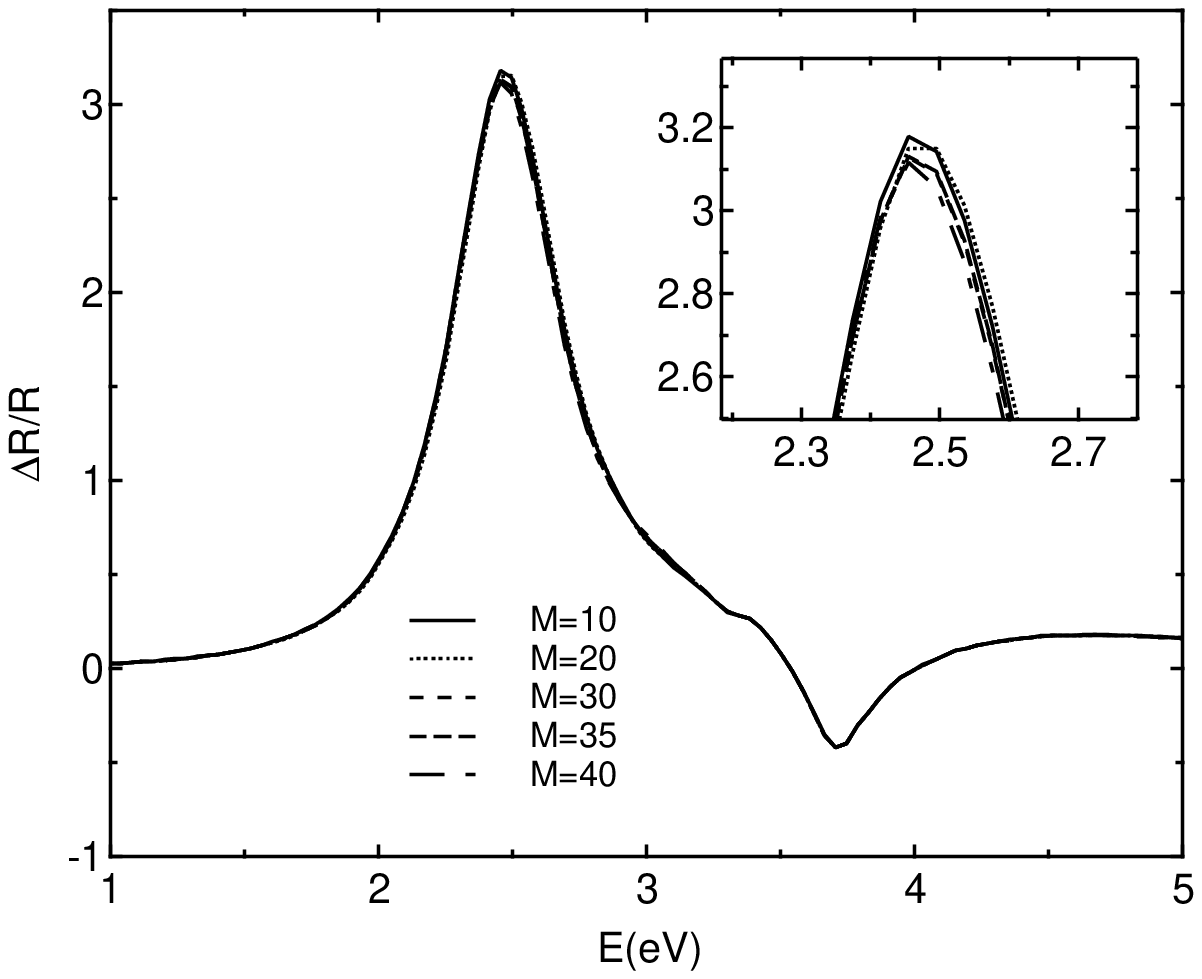,width=12cm}
  \end{center}
    \mycaption{\myauthor}{\mytitle}
\end{figure}

\newpage

\begin{figure}[h]
  \begin{center}
    \epsfig{file=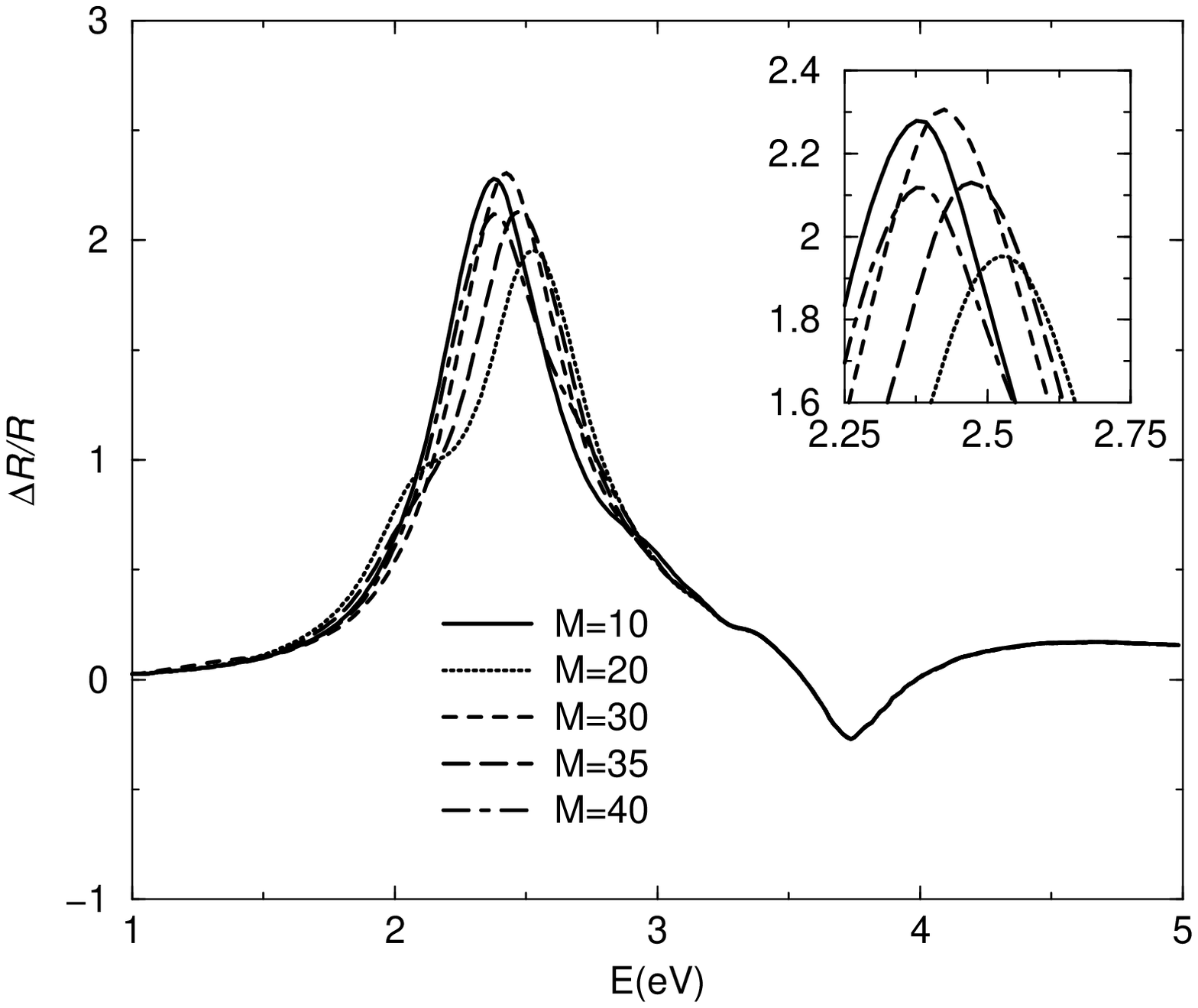,width=12cm}
  \end{center}
    \mycaption{\myauthor}{\mytitle}
\end{figure}

\newpage

\begin{figure}[h]
  \begin{center}
    \epsfig{file=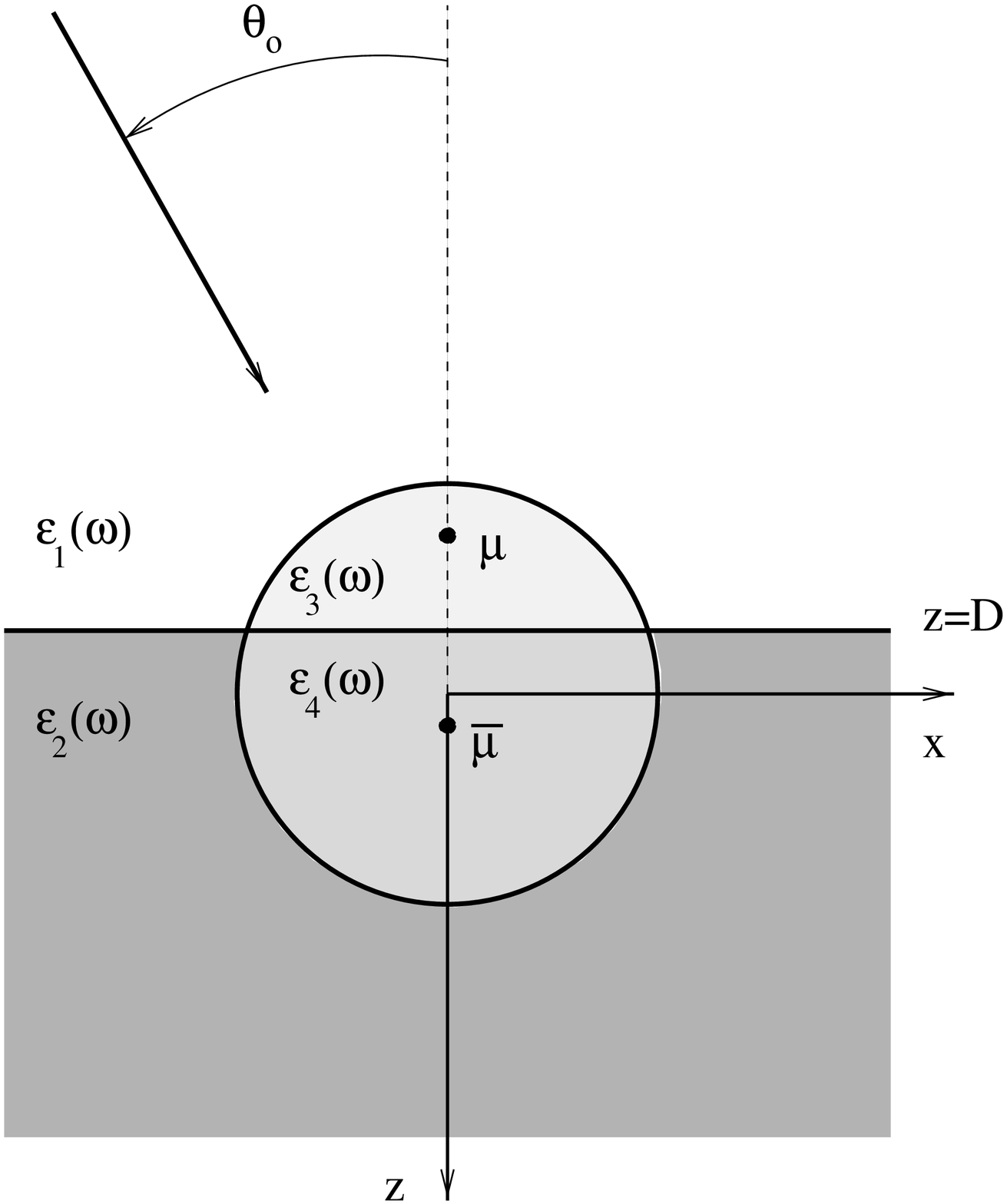,width=12cm}
  \end{center}
    \mycaption{\myauthor}{\mytitle}
\end{figure}

\newpage

\begin{figure}[h]
  \begin{table*}
    \begin{center}
      \epsfig{file=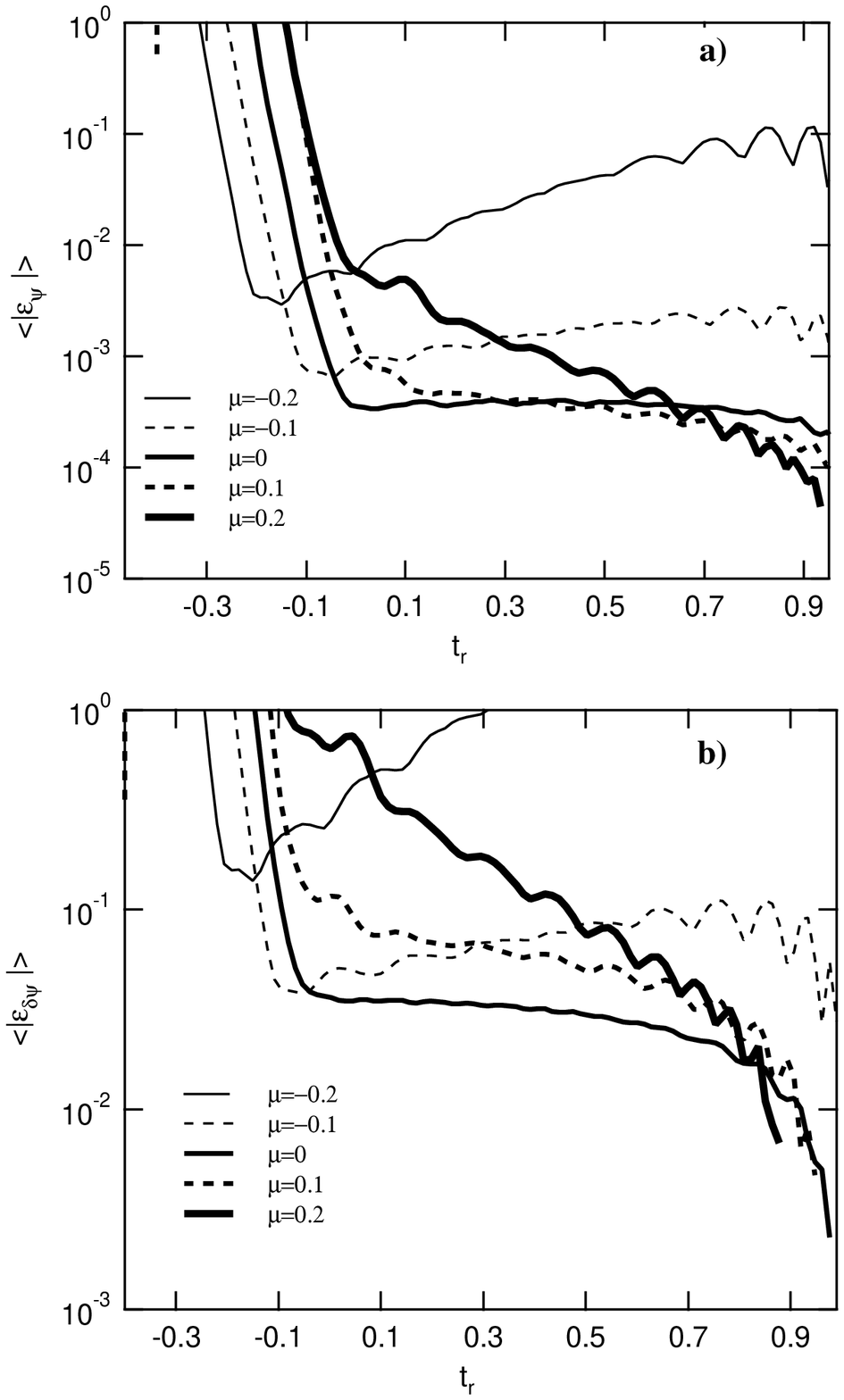,width=14cm} 
    \end{center}
  \mycaption{\myauthor}{\mytitle}
  \end{table*}
\end{figure}

\newpage

\begin{figure}[h]
  \begin{center}
    \epsfig{file=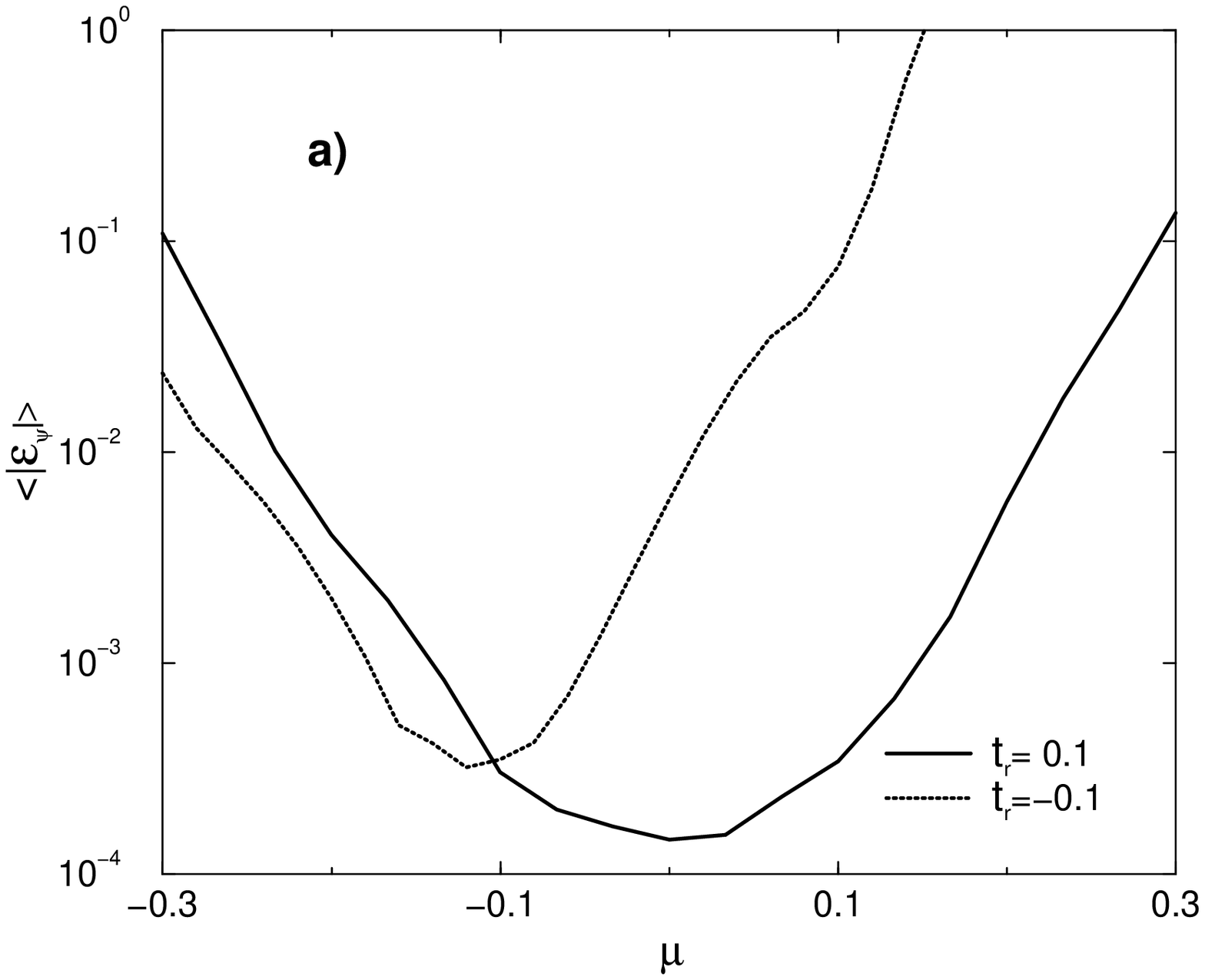,width=12cm}\\*[0.7cm]
    \epsfig{file=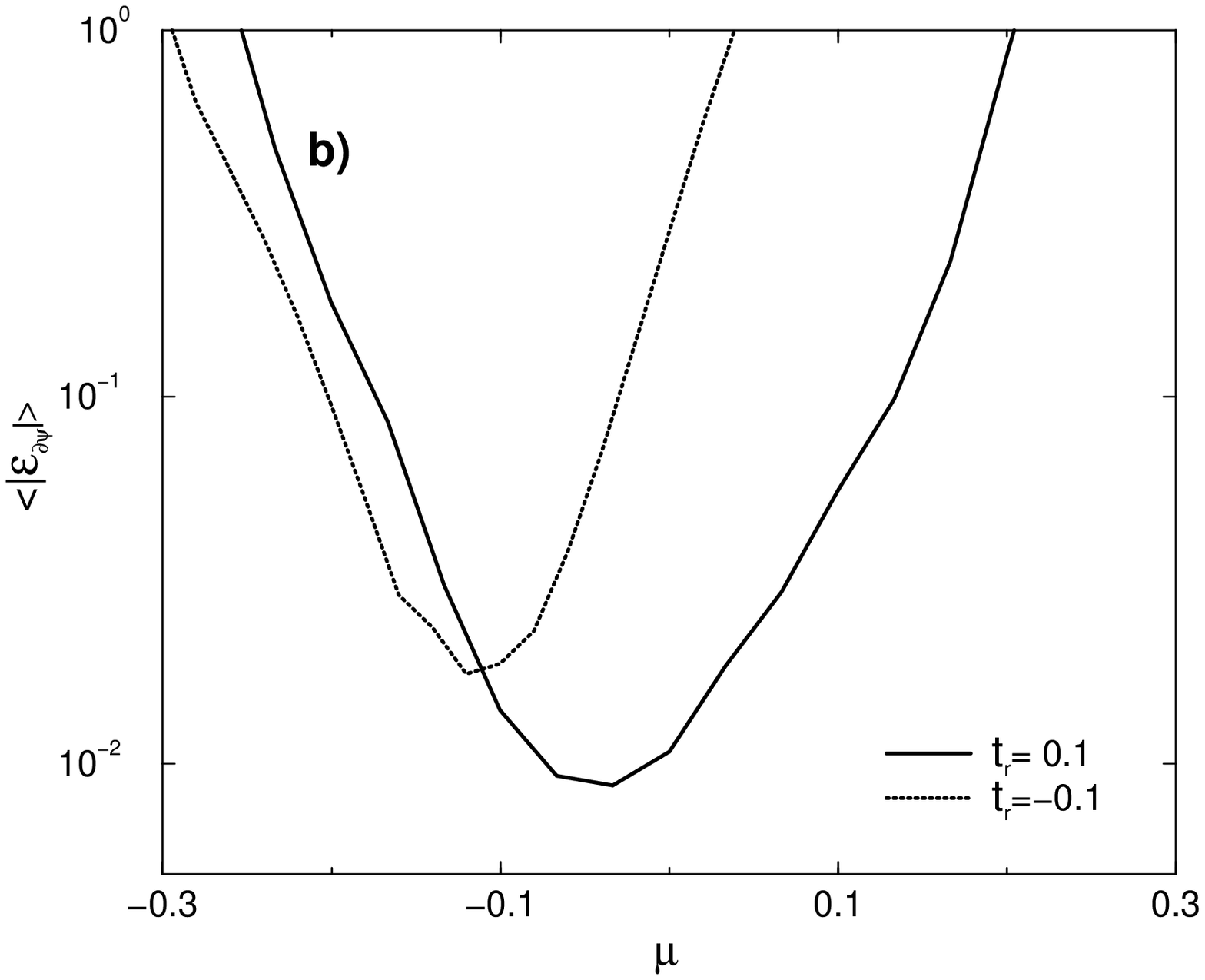,width=12cm}  
  \end{center}
    \mycaption{\myauthor}{\mytitle}
\end{figure}

\newpage

\begin{figure}[h]
  \begin{table*}
    \begin{center}
      \epsfig{file=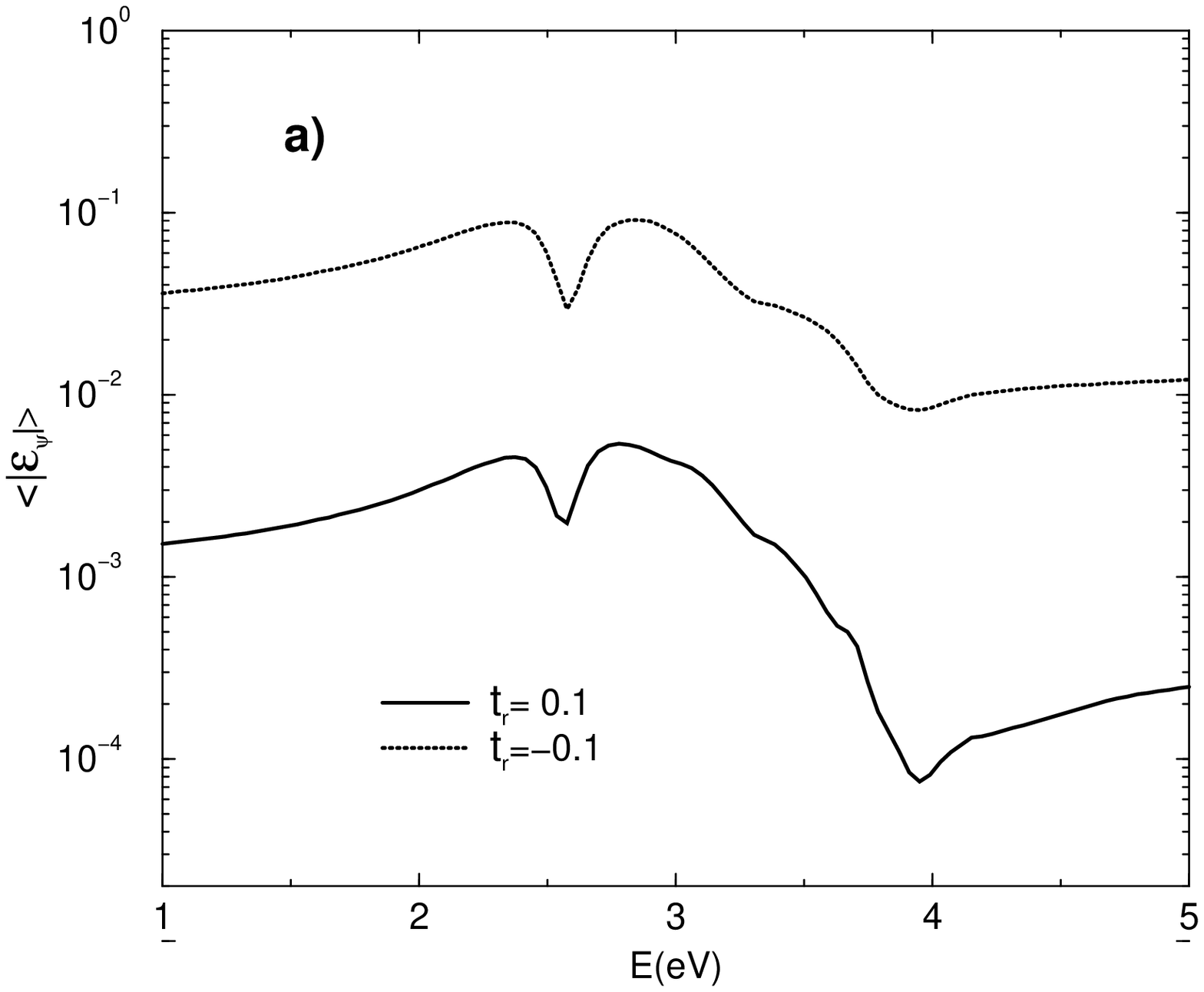,width=12cm} \\*[0.7cm]
      \epsfig{file=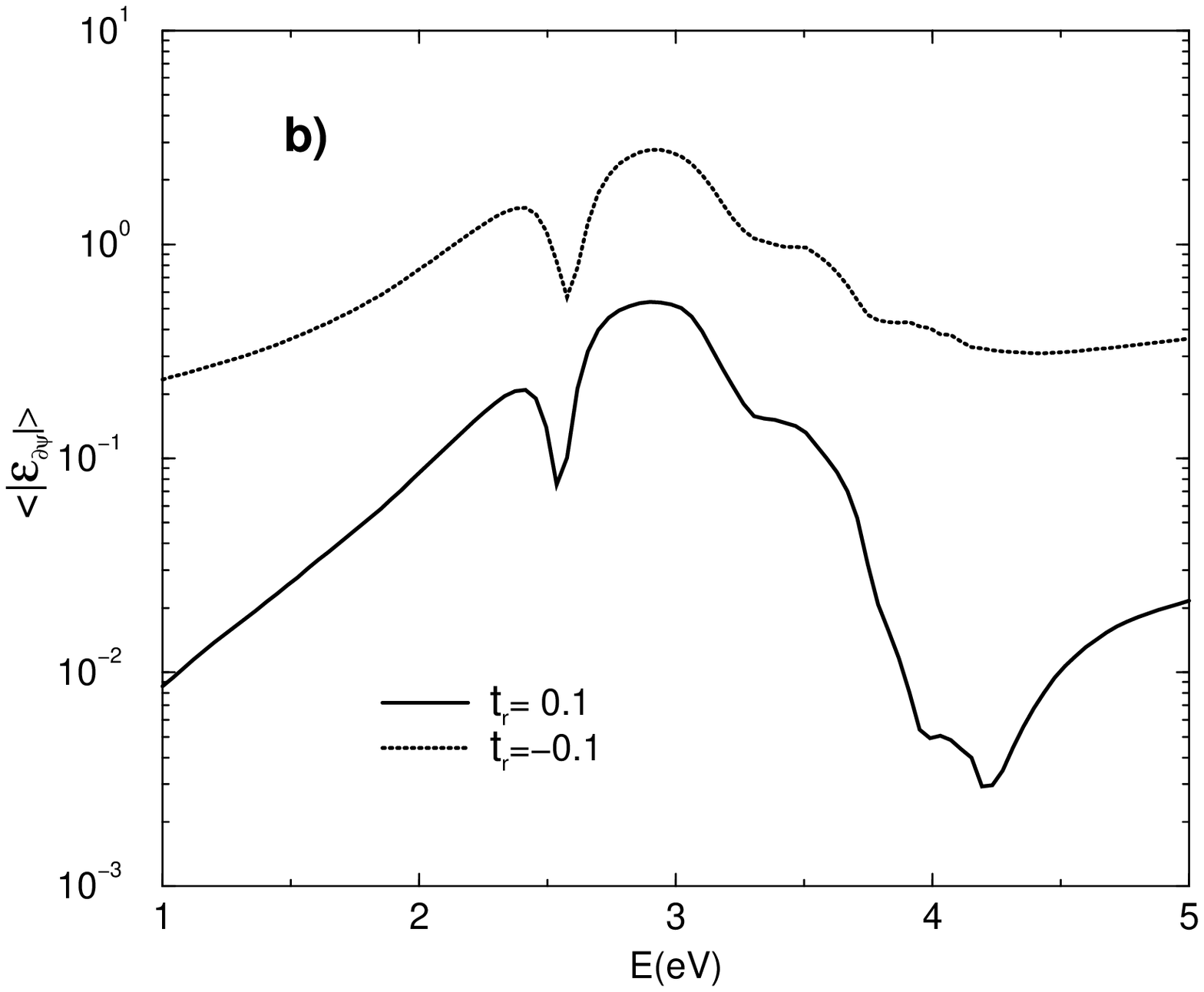,width=12cm}  
    \end{center}
  \mycaption{\myauthor}{\mytitle}
  \end{table*}
\end{figure}

\newpage

\begin{figure}[h]
  \begin{table*}
    \begin{center}
      \epsfig{file=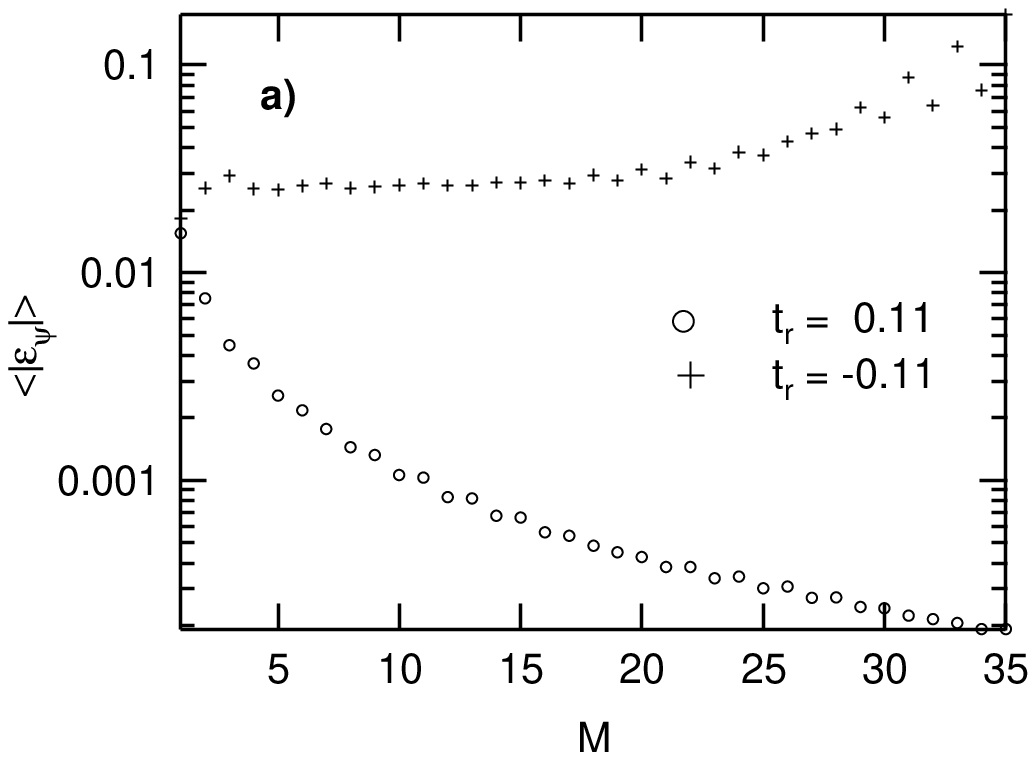,width=12cm} \\*[0.7cm]
      \epsfig{file=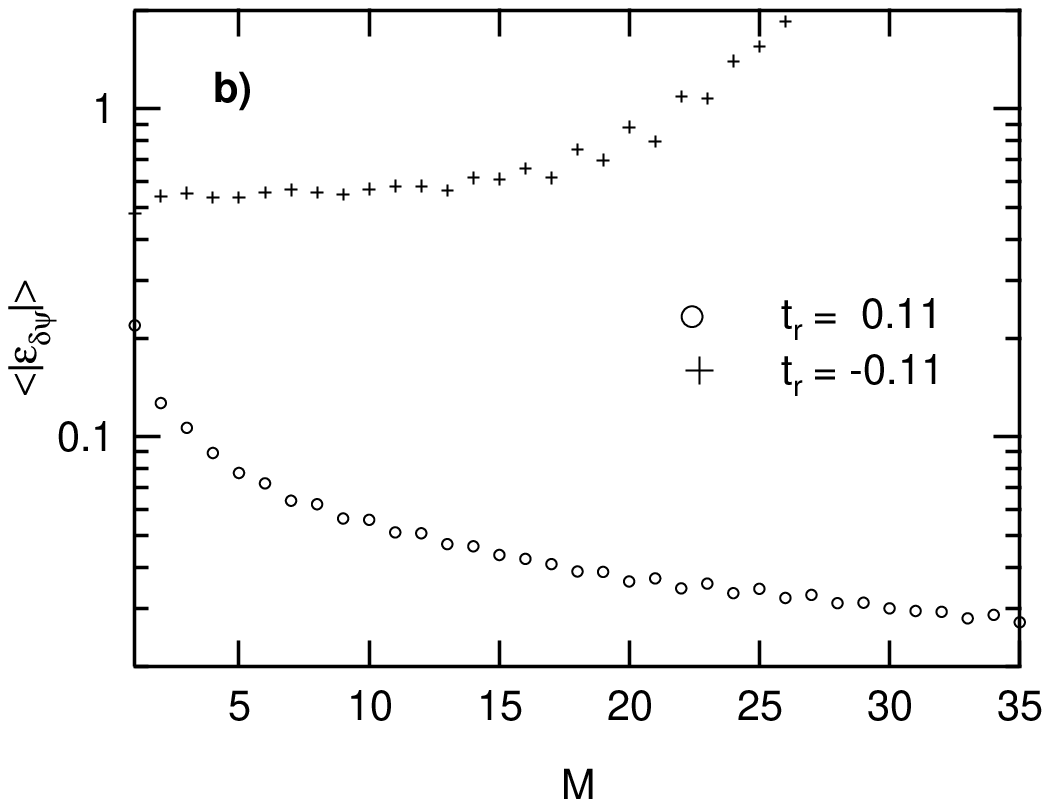,width=12cm}  
    \end{center}
  \mycaption{\myauthor}{\mytitle}
  \end{table*}
\end{figure}

\newpage

\begin{figure}[h]
  \begin{center}
    \epsfig{file=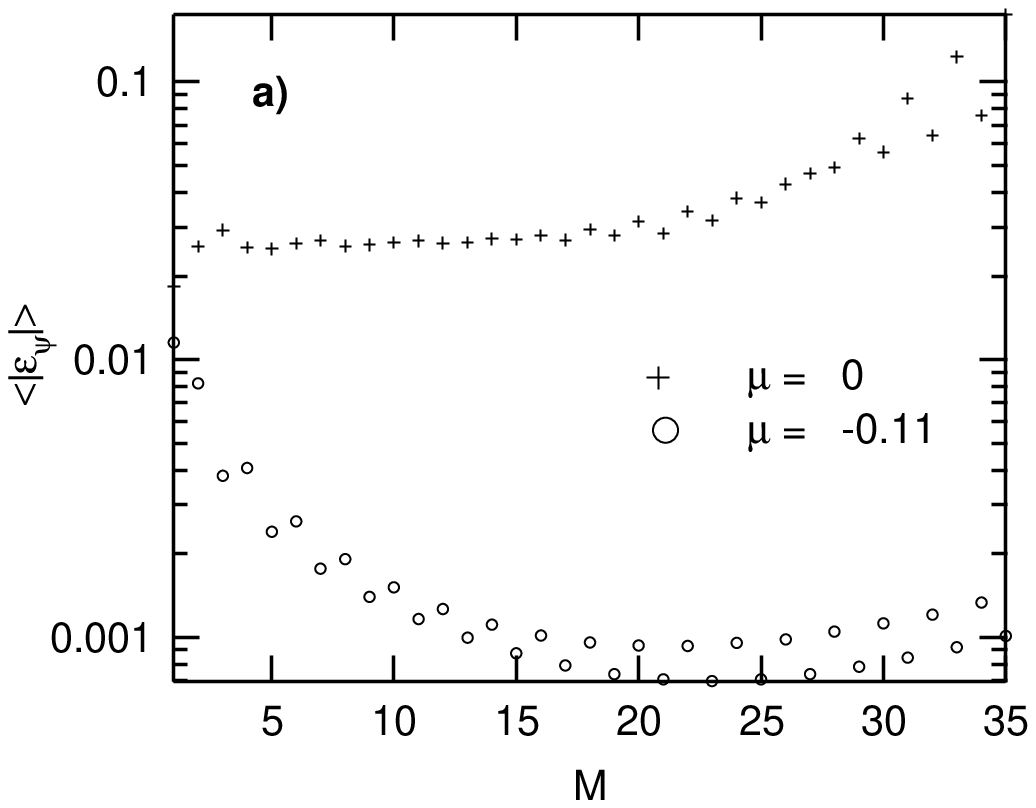,width=12cm}\\*[0.7cm]
    \epsfig{file=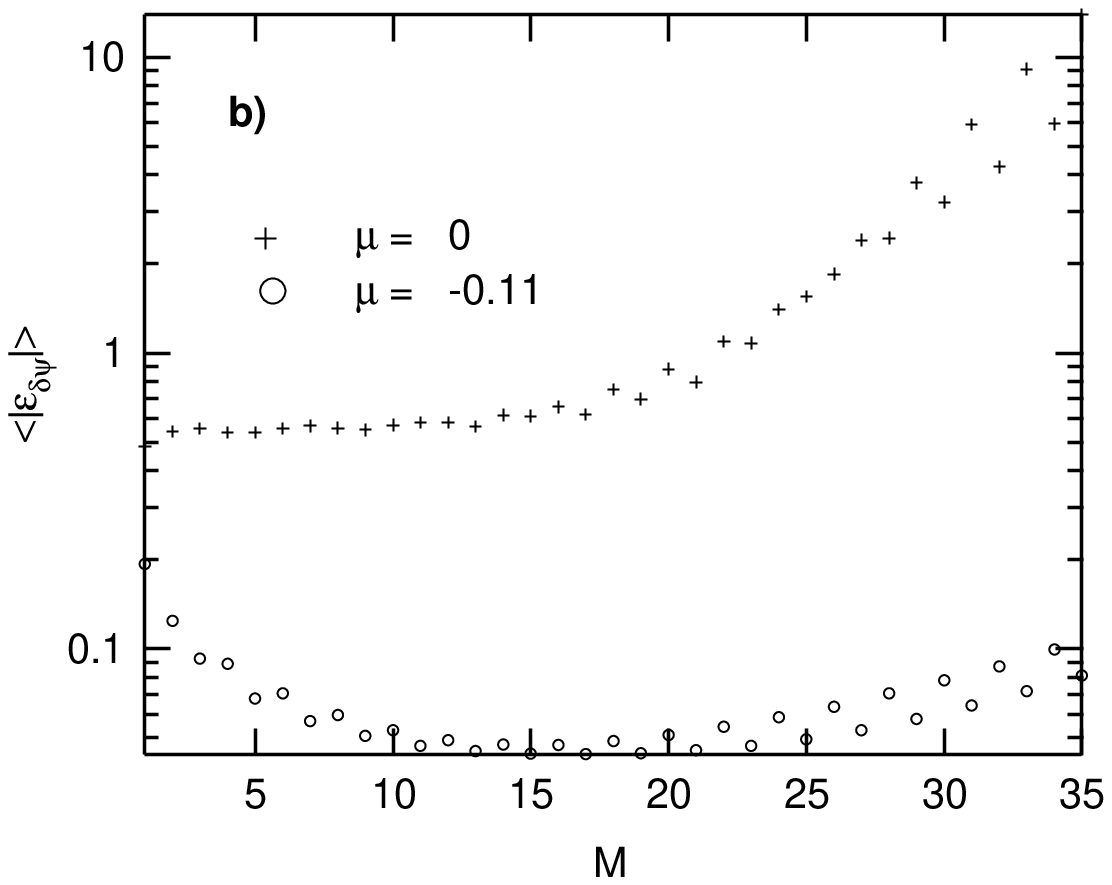,width=12cm} 
  \end{center}
    \mycaption{\myauthor}{\mytitle}
\end{figure}

\newpage

\begin{figure}[h]
  \begin{center}
    \epsfig{file=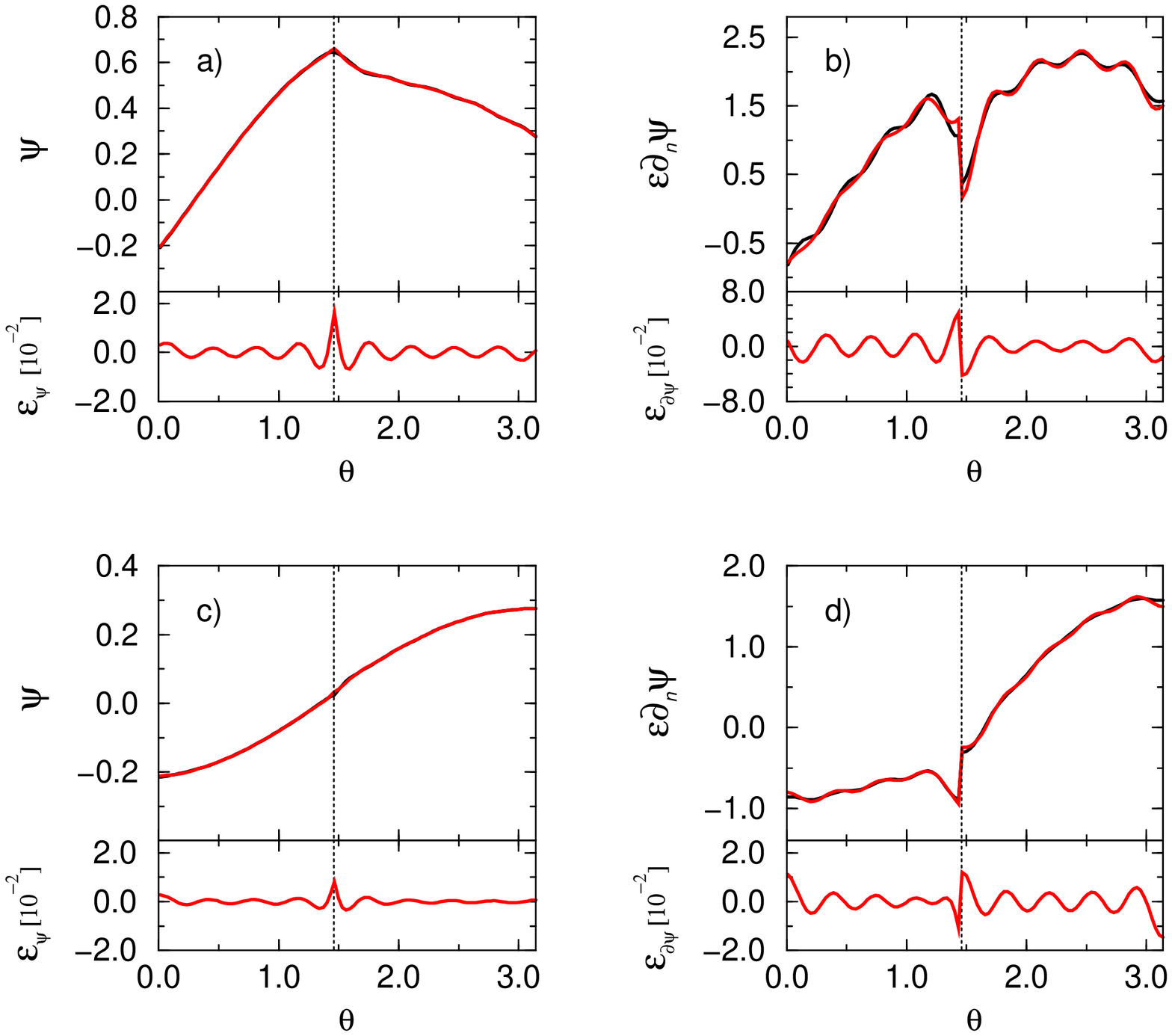,width=16cm}
  \end{center}
    \mycaption{\myauthor}{\mytitle}
\end{figure}

\newpage

\begin{figure}[h]
  \begin{center}
    \Large a) \epsfig{file=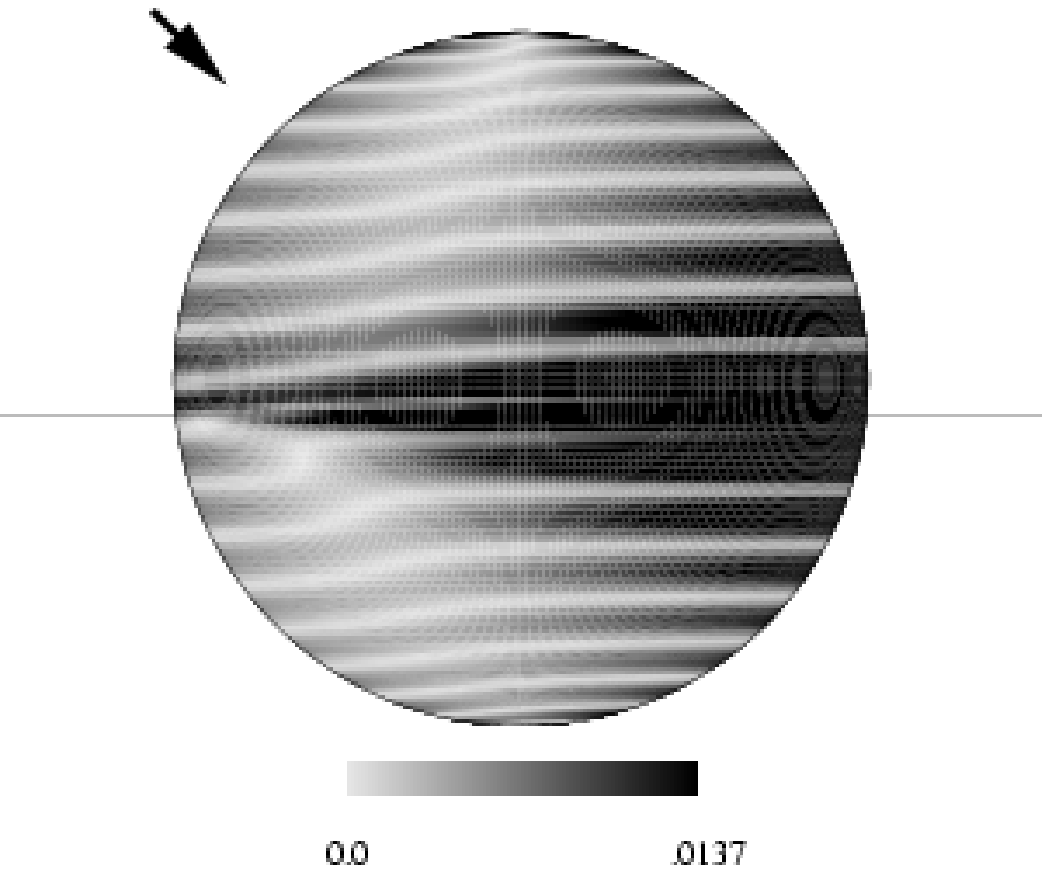,width=10cm}\\*[0.7cm]
    \Large b) \epsfig{file=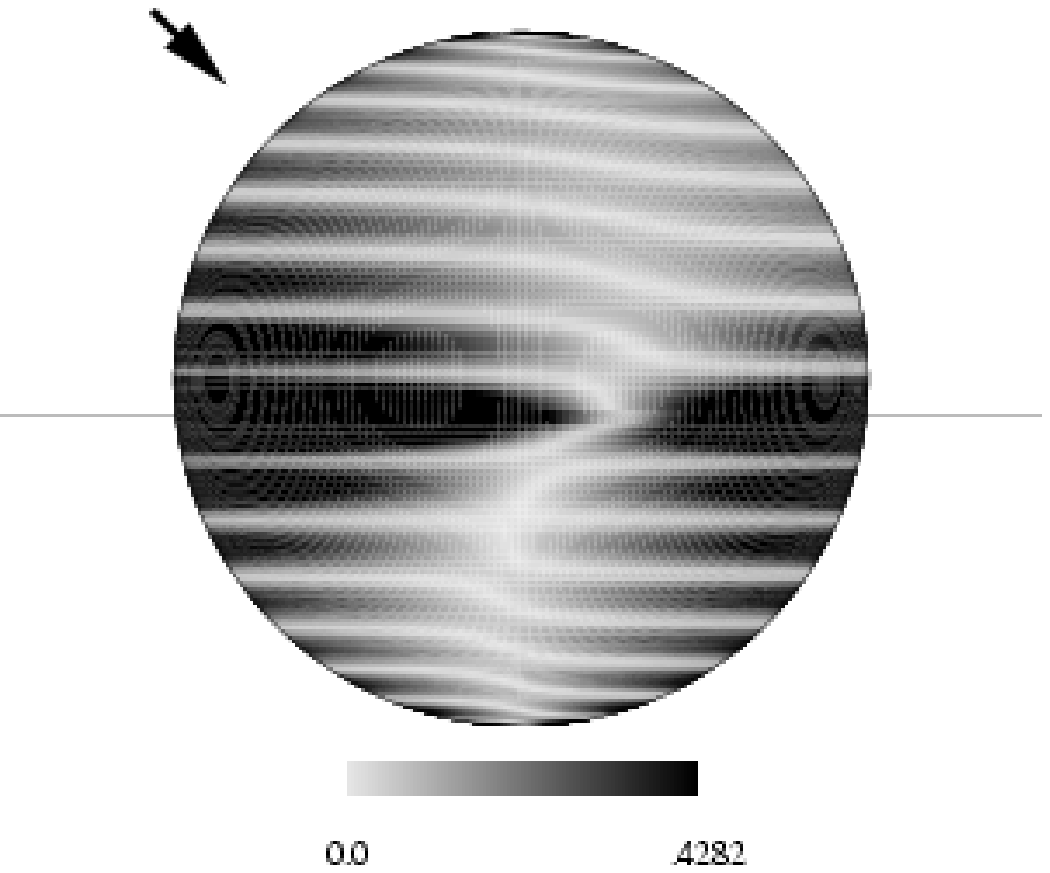,width=10cm} 
  \end{center}
    \mycaption{\myauthor}{\mytitle}
\end{figure}


\begin{thebibliography}{99}
\bibitem{Wind87a} 
   M.\ M.\ Wind, J.\ Vlieger, D.\ Bedeaux, Physica A {\bf 141}, 33 (1987). 

\bibitem{Wind87b} 
   M.\ M.\ Wind, P.\ A.\ Bobbert, J.\ Vlieger, D.\ Bedeaux, 
      Physica A {\bf 143}, 164 (1987). 

\bibitem{Renaud_SSReports}
   G.\ Renaud, Surface Science Reports {\bf 32}, 1/2 (1998).

\bibitem{Aspnes}
    D.E.\ Aspnes, N.\ Dietz, Appl.\ Surf.\ Sci.\  {\bf 130-132}, 367 (1998).

\bibitem{Yves1}
    Y.\ Borensztein, Physica A {\bf 207}, 293 (1994).

\bibitem{Doro1}
        D.\ Martin, F.\ Creuzet, J.\ Jupille, Y.\ Borensztein, P.\
        Gadenne, Surf.\ Sci.\ {\bf 377-379}, 958 (1997).

\bibitem{Doro2}         
        D.\ Martin, J.\ Jupille, Y.\ Borensztein,
         Surf.\ Sci.\ {\bf 402-404}, 433 (1998).
    

\bibitem{Doro3}
        D.\ Martin, J.\ Jupille, unpublished results.
        

\bibitem{Remi}
        R.\ Lazzari, J. Jupille, Y.\ Borensztein, Appl.\ Surf.\ Sci.\ {\bf 142}, 451 (1999).

\bibitem{Maxwell-Garnett}
       J.\ C.\ Maxwell Garnett, Phil.\ Trans.\ Roy.\ Soc.\ London {\bf 203A},
       385 (1904).
  

\bibitem{KreibigBook}
        U.\ Kreibig, M.\ Vollmer, 
        {\it Optical Properties of Metal Clusters, Springer Series in
        Material Science} {\bf 25}, Berlin (1995).

\bibitem{Mie}
        G.\ Mie, Ann.\ Phys.\ {\bf 25}, 377 (1908).

\bibitem{Landauer} 
    R.\ Landauer in {\it Proceedings of the First Conference on the
    Electromagnetic and Optical Properties of Inhomogeneous Media},
    eds. J.\ C.\ Garland and D.\ B.\ Tanner, New York (1978).  

\bibitem{Barrera2}
        R.\ G.\ Barrera, M.\ del Castillo-Mussot, G.\ Monsivais, 
        P.\ Villase\~nor, W.\ L.\ Moch\'an, Phys.\ Rev.\ B {\bf 43}, 13819 (1991). 

\bibitem{Barrera1}
        R.\ G.\ Barrera, G.\ Monsivais, L.\ Moch\'an, Phys. Rev. B {\bf 38}, 5371 (1988).

\bibitem{Yamaguchi1}
        T.\ Yamaguchi, S.\ Yoshida, A.\ Kinbara, Thin Solid Films {\bf 18}, 63 (1973).

\bibitem{Yamaguchi2}
        T.\ Yamaguchi, S.\ Yoshida, A.\ Kinbara, Thin Solid Films {\bf 21}, 173 (1974).

\bibitem{Bedeaux73} 
        D.\ Bedeaux, J.\ Vlieger, Physica A {\bf 67}, 55 (1973).  
 

\bibitem{Bedeaux74}
        D.\ Bedeaux, J.\ Vlieger, Physica A {\bf 73}, 287 (1974).

\bibitem{Bedeaux76}
        J.\ Vlieger, D.\ Bedeaux, Physica A {\bf 82}, 221 (1976).

\bibitem{Bedeaux80}
        D.\ Bedeaux, J.\ Vlieger, Thin Solid Films {\bf 69}, 107 (1980).

\bibitem{Bedeaux83}
       J.\ Vlieger, D.\ Bedeaux, Thin Solid Films {\bf 102}, 265 (1983).

\bibitem{Kretschmann} 
    E.\ Kretschmann, Z.\ Phys.\ {\bf 227}, 412 (1969).  

\bibitem{Jackson}
        J.\ D.\ Jackson, {\it Classical Electrodynamics}, 
         John Wiley and Sons Editions New York, (1975).

\bibitem{Bedeaux88}
    M.\ M.\ Wind, P.\ A.\ Bobbert, J.\ Vlieger, Thin Solid Films {\bf 164}, 57 (1988).

\bibitem{Fabrice} 
    F.\ Didier and J.\ Jupille, J. Adh. {\bf 58}, 253 (1996).

\bibitem{Haarmans93}
        M.\ T.\ Haarmans, D.\ Bedeaux, Thin Solid Films {\bf 224}, 117 (1993).

\bibitem{Palik}
        E.\ D.\ Palik, {\it Handbook of Optical Constants of Solids}, Academic Press, New York, (1985).

\bibitem{Morse} 
    P.\ M.\ Morse and H.\ Feshbach, 
       {\it Methods of Theoretical Physics}, Part 1 and 2, McGraw-Hill, New-York, (1953).








\end{thebibliography}
\end{document}